\documentclass{aa}  

\usepackage{graphicx}
\usepackage{txfonts}
\usepackage[]{hyperref}
\usepackage{amsmath}
\usepackage{multirow}
\usepackage{enumitem}
\usepackage{color}
\usepackage{ulem}
\usepackage{CJKfntef,CJK}
\usepackage{makecell}

\newcommand\kms{$\rm km\ s^{-1}$}
\newcommand\coa{$\rm ^{12}CO$}
\newcommand\hcop{$\rm HCO^+$}
\newcommand\hcn{$\rm HCN$}
\newcommand\cch{$\rm C_2H$}
\newcommand\hhco{$\rm H_2CO$}
\newcommand\occthreehtwo{\textit{o-c-}$\rm C_3H_2$}
\newcommand\ccthreehtwo{\textit{c-}$\rm C_3H_2$}
\newcommand\hcoponco{$N({\rm HCO^+})/N({\rm CO})$}
\newcommand\hcnonco{$N({\rm HCN})/N({\rm CO})$}
\newcommand\cchonco{$N({\rm C_2H})/N({\rm CO})$}
\newcommand\octhtonco{$N({o\text{-}c\text{-}\rm C_3H_2})/N({\rm CO})$}
\newcommand\cthtonco{$N({c\text{-}\rm C_3H_2})/N({\rm CO})$}
\newcommand\conco{$N({\rm C^0})/N({\rm CO})$}

\begin{document}

   \title{Molecular chemistry induced by J-shock toward supernova remnant W51C}


   \author{Tian-Yu Tu\inst{1}
          \and
          Valentine Wakelam\inst{2}
          \and
          Yang Chen\inst{1,3}
          \and
          Ping Zhou\inst{1,3}
          \and
          Qian-Qian Zhang\inst{1}
          }

   \institute{
             School of Astronomy \& Space Science, Nanjing University, 163 Xianlin Avenue, Nanjing 210023, China\\
             \email{ygchen@nju.edu.cn}
         \and
             Laboratoire d'Astrophysique de Bordeaux, Univ. Bordeaux, CNRS, B18N, allée Geoffroy Saint-Hilaire, 33615 Pessac, France
         \and
             Key Laboratory of Modern Astronomy and Astrophysics, Nanjing University, Ministry of Education, Nanjing 210023, China
             }

   \date{Received ...; accepted ...}

 
  \abstract
   {Shock waves from supernova remnants (SNRs) have strong influence on the physical and chemical properties of molecular clouds (MCs). 
   Shocks propagating into magnetized MCs can be classified into ``jump'' J-shock and ``continuous'' C-shock. 
   The molecular chemistry in the re-formed molecular gas behind J-shock is still not well understood, which will provide a comprehensive view of the chemical feedback of SNRs and the chemical effects of J-shock. }
   {We conducted a W-band (71.4--89.7 GHz) observation toward a re-formed molecular clump behind a J-shock induced by SNR W51C with the Yebes 40 m radio telescope to study the molecular chemistry in the re-formed molecular gas.  
   }
   {Based on the local thermodynamic equilibrium (LTE) assumption, we estimate the column densities of \hcop, \hcn, \cch\ and \occthreehtwo, and derive the maps of their abundance ratios with CO. 
   The gas density is constrained by non-LTE analysis of the \hcop\ $J=1$--0 line. 
   The abundance ratios are compared with the values in typical quiescent MCs and shocked MCs, and are also compared with the results of chemical simulations with the Paris-Durham shock code to verify and investigate the chemical effects of J-shock. 
   }
   {We obtain the following abundance ratios: $N({\rm HCO^+})/N({\rm CO})\sim (1.0\text{--}4.0)\times 10^{-4}$, $N({\rm HCN})/N({\rm CO})\sim (1.8\text{--}5.3)\times 10^{-4}$, $N({\rm C_2H})/N({\rm CO})\sim (1.6\text{--}5.0)\times 10^{-3}$, and $N({o\text{-}c\text{-}{\rm C_3H_2}})/N({\rm CO})\sim (1.2\text{--}7.9)\times 10^{-4}$. 
   The non-LTE analysis suggests that the gas density is $n_{\rm H_2}\gtrsim 10^4\rm \ cm^{-3}$. 
   We find that the \cchonco\ and \octhtonco\ are higher than typical values in quiescent MCs and shocked MCs by 1--2 orders of magnitude, which can be qualitatively attributed to the abundant $\rm C^+$ and C at the earliest phase of molecular gas re-formation. 
   The Paris-Durham shock code can reproduce, although not perfectly, the observed abundance ratios, especially the enhanced \cchonco\ and \cthtonco, with J-shocks propagating in to both non-irradiated and irradiated molecular gas with a preshock density of $n_{\rm H}=2\times 10^3\rm \ cm^{-3}$. }
   {}

   \keywords{shock waves --
             ISM: molecules --
             ISM: clouds --
             ISM: supernova remnants --
             ISM: individual objects: W51C
               }

   \maketitle
%

\section{Introduction}
Supernova remnants (SNRs) exert strong influence on the physical and chemical properties of the molecular clouds (MCs) they interact with \citep[e.g.,][]{vanDishoeck_Submillimeter_1993}. 
Among all the effects driven by SNRs, the shock waves play a crucial rule. 
Shock can heat, compress and accelerate the molecular gas \citep{Draine_Theory_1993}, altering the physical properties of the MC, and regulate the molecular chemistry through various processes \citep[e.g.,][]{Burkhardt_Modeling_2019}. 

\par

Shocks propagating into magnetized MCs can be roughly classified into two types: J-shocks and C-shocks \citep{Draine_Theory_1993}. 
J-shocks, where J refers to ``jump'', are often fast and weakly magnetized. 
The deceleration and heating of the entire shock is almost entirely due to the viscous stresses arising in a thin transition layer called the viscous subshock. 
A jump of physical parameters (density, temperature, etc.) is expected in the shock profile.
On the contrary, C-shocks, where C refers to ``continuous'', are often slow and strongly magnetized. 
The momentum transfer and heating are finished by the magnetic precursor which goes ahead of the viscous subshock when the magnetic field is strong enough to make the magnetosonic speed ($B/\sqrt{4\pi\rho_i}$) higher than the shock velocity. 
Therefore, the physical parameters change continuously without a jump between the upstream and downstream.

\par

Molecular chemistry induced by C-shocks has been extensively studied in SNRs \citep[e.g.,][]{vanDishoeck_Submillimeter_1993,Lazendic_Multiwavelength_2010,Zhou_Unusually_2022b,Tu_Shock_2024} and other astrophysical environments such as molecular outflows of protostars \citep[e.g.,][]{Bachiller_Chemically_2001,Mendoza_search_2018,Codella_Seeds_2020}. 
Since C-shocks are non-dissociative, most molecules can survive, and C-shocks can also release the depleted molecular species back to the gas phase and make them detectable \citep[e.g., SiO,][]{Gusdorf_SiO_2008}. 
However, investigations of molecular chemistry induced by J-shock are hampered by the fact that J-shock can result in much higher temperature than C-shock and most molecules will be dissociated \citep[e.g.,][]{Kristensen_Shock_2023a}. 
Therefore, observational studies of J-shock often turn to emission lines of atoms \citep[e.g.,][]{Rho_Near-Infrared_2001,Lee_infrared_2019} or infrared transitions of a limited variety of molecules: CO, $\rm H_2$ and $\rm H_2O$ \citep[e.g.,][]{Shinn_Ortho-to-para_2012,Rho_Detection_2015}. 
However, simulations have found that molecules can re-form in the cooling gas behind a fast and dissociative J-shock \citep{Hollenbach_Molecule_1989,Neufeld_Fast_1989,Cuppen_H2_2010,Hollenbach_Interstellar_2013}, and observations have also found this re-formed molecular gas in SNRs W51C \citep{Koo_Interaction_1997} and IC443 \citep{Wang_Strongly_1992}.
In the prototypical protostellar outflow L1157, molecular gas re-formed behind J-shock was identified by CO observations \citep{Lefloch_CHESS_2012,Benedettini_CHESS_2012}, and re-formed CS was discovered \citep{Gomez-Ruiz_density_2015}. 
However, the abundance of CS deviates little from other C-shocked components. 
It is still unclear whether J-shock can induce different chemistry compared with C-shock and quiescent gas conditions. 

\par

SNR W51C (G49.2$-$0.7) is a middle-aged SNR interacting with MCs, evidenced by 1720 MHz OH masers \citep{Green_Continuation_1997}, broadened molecular line \citep{Koo_Interaction_1997,Brogan_OH_2013}, SiO emission \citep{Dumas_Localized_2014}, etc.  
The cold CO gas re-formed behind a J-shock was found by \citet{Koo_Interaction_1997} based on three observational facts. 
First, the local-standard-of-rest (LSR) velocity of the CO gas ($\gtrsim 80$ \kms) is larger than the tangent point LSR velocity ($\approx 60$ \kms) toward this direction ($l\approx49^\circ$), so it is likely that the CO gas is accelerated by the shock to high LSR velocity. 
Second, the line-of-sight velocity separation between the preshock and postshock CO gas is large ($\approx20$--50 \kms), while for C-shock, the postshock gas should be at velocities closer to that of the preshock gas since the acceleration of the shocked gas is continuous. 
Third, the estimated excitation temperature of the CO emission is low ($\sim 10\rm \ K$), which should be higher for C-shocked gas. 
Fourth, the postshock CO gas exhibits spatial and spectral coincidence with high-velocity HI gas which has been proposed to be subject to a fast shock with a velocity of $\sim 100$ \kms\ \citep{Koo_Interaction_1997a}. 

\par

In this work, we present our new molecular observation toward the ``clump 2'' of W51C found and named by \citet{Koo_Interaction_1997} to study the detailed molecular chemistry induced by J-shock. 
In Section \ref{sec:obs}, we describe the details about our observation and data reduction. 
The observational results are presented in Section \ref{sec:res}. 
In Section \ref{sec:dis}, we calculate the molecular column densities and abundance ratios, constrain the gas density, discuss the chemistry of carbon chain species, and present the results of chemical simulation. 
Our conclusions are summarized in Section \ref{sec:con}. 

\section{Observations} \label{sec:obs}

\begin{figure}
\centering
\includegraphics[width=0.95\hsize]{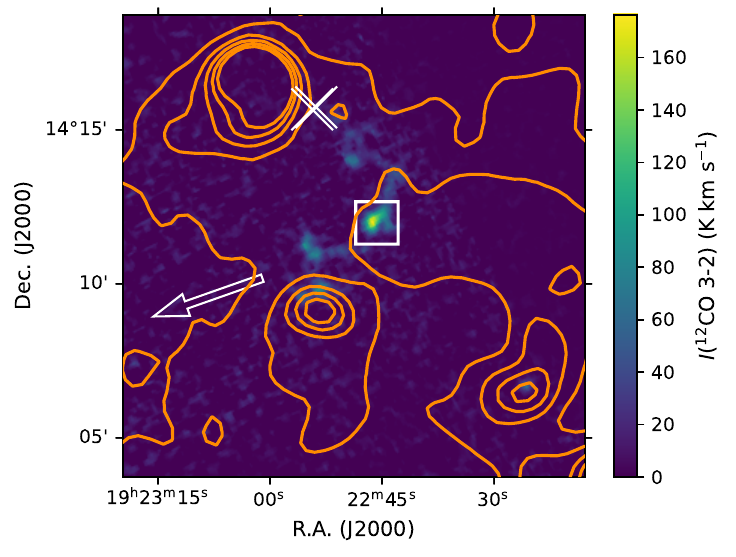}
\caption{Integrated intensity map of the JCMT \coa\ $J=3$--2 line in velocity range $+80$--$+110$ \kms, which is the velocity range of the re-formed CO gas proposed by \citet{Koo_Interaction_1997}, overlaid with orange contours of VGPS 1.4 GHz radio continuum (levels are 50--250 K in steps of 50 K). 
The two white crosses shows the 1720 MHz OH masers discovered by \citet{Green_Continuation_1997}. 
The white arrow points to the geometric center of W51C recorded in the Green's SNR catalog \citep{Green_revised_2019}. 
The white box shows the molecular clump toward which we conducted the Yebes 40 m observation. 
}
\label{fig1}
\end{figure}

\subsection{Yebes 40 m observation}
We conducted new mapping observation toward clump 2 (which was identified and named by \citet{Koo_Interaction_1997}) of SNR W51C with the Yebes 40 m radio telescope (PI: Tian-Yu Tu, project code: 24A003) in W-band covering a spectral range of 71.4--89.7 GHz. 
Position switching mode was adopted throughout the observation with the reference point at $\alpha_{J2000}={\rm 19^h23^m05^s.8}$, $\delta_{J2000}=+14^\circ10^\prime56^{\prime\prime}$. 
The mapping was made toward a $82^{\prime\prime}.5\times 82^{\prime\prime}.5$ region centered at $\alpha_{J2000}={\rm 19^h22^m45^s.69}$, $\delta_{J2000}=+14^\circ11^\prime58^{\prime\prime}.9$ in a pixel size of $7^{\prime\prime}.5\times 7^{\prime\prime}.5$ (see Figure \ref{fig1} for the position and size of the mapped region). 
The sensitivity measured in main beam temperature $T_{\rm mb}$ at the raw spectral resolution (38 kHz) is 0.05--0.09 K. 
The data was reduced with the GILDAS/CLASS package\footnote{\url{https://www.iram.fr/IRAMFR/GILDAS/}}. 
For better comparison, all of the reduced data cubes are smoothed to a common beam size of $27^{\prime\prime}.6$.  

\subsection{Other archival data} \label{sec:data2}
We obtained some archival data to support our analysis. 
The \coa\ $J=3$--2 data was retrieved from the \coa\ (3–2) High-Resolution Survey (COHRS) project \citep{Park_12CO_2023} performed by the James Clarke Maxwell Telescope (JCMT). 
The angular resolution is $16.6^{\prime\prime}$ and the sensitivity measured in $T_{\rm A}^*$ is $\sim 1\rm \ K$ at a velocity channel width of 0.635 \kms. 
The antenna temperature was converted to $T_{\rm mb}$ with a main beam efficiency of 0.61. 

\par

We also obtained the \coa\ $J=1$--0 data observed by the Nobeyama 45\,m telescope and the [C\,I] $({^3P_1}\text{--}{^3P_0})$ data observed by the Atacama Submillimeter Telescope Experiment (ASTE) 10\,m telescope from \citet{Yamagishi_Cosmic-ray-driven_2023}. 
Details of the observation can be found therein. 
All of these supplementary data cubes are smoothed to a beam size of $27^{\prime\prime}.6$. 

\par

The 1.4 GHz radio continuum map of W51C was obtained from the VLA Galactic Plane Survey \citep[VGPS,][]{Stil_VLA_2006}. 
All the processed data were further analyzed with \textit{Python} packages Astropy \citep{AstropyCollaboration_Astropy_2018,AstropyCollaboration_Astropy_2022} and Spectral-cube \footnote{\url{https://spectral-cube.readthedocs.io/en/latest/}}.
The data cubes of the CO isotopes were reprojected with the reproject\footnote{\url{https://reproject.readthedocs.io/en/stable/}} package. 
We visualized the data with \textit{Python} package Matplotlib\footnote{\url{https://matplotlib.org/}}.

\section{Results} \label{sec:res}

\begin{figure*}
\centering
\includegraphics[width=0.9\textwidth]{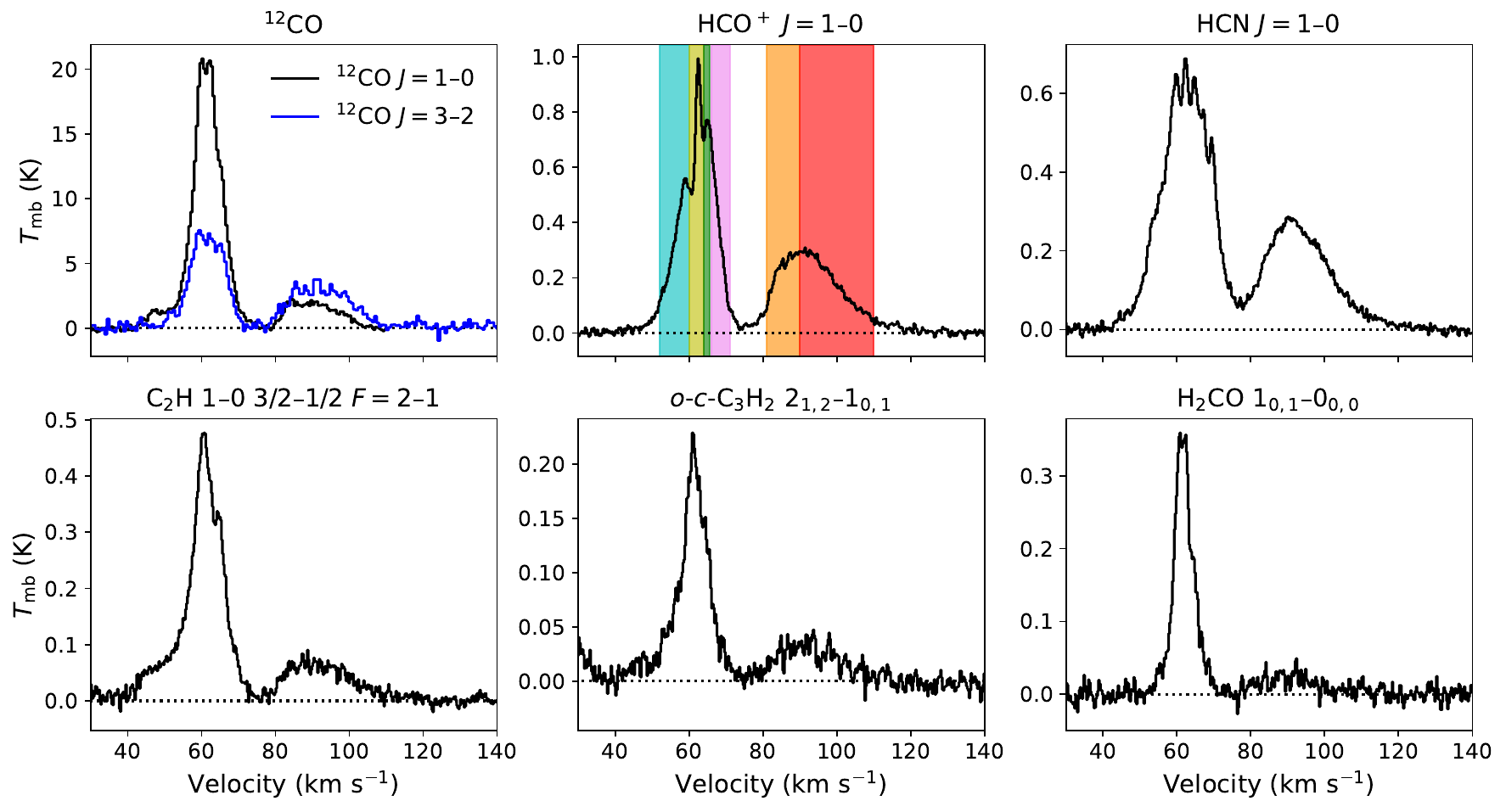}
\caption{Spectra of the molecular transitions with emission detected in velocity range $+80$--$+110$ \kms\ averaged across the entire mapped region. 
Velocity integration intervals for the six velocity components are shown by colored rectangles in the spectrum of \hcop. 
}
\label{fig:spec}
\end{figure*}

\begin{figure*}
\centering
\includegraphics[width=0.9\textwidth]{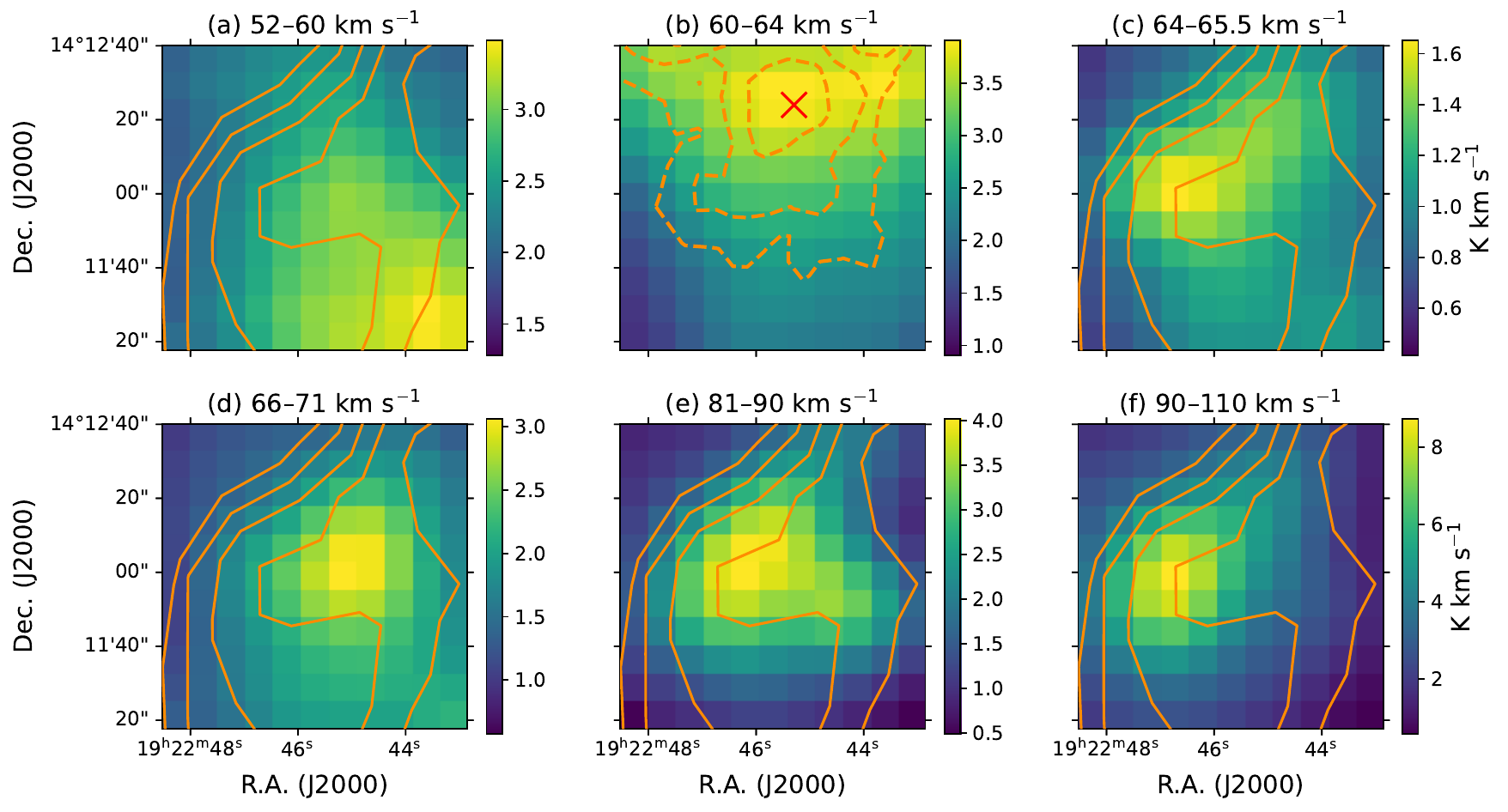}
\caption{\hcop\ Integrated intensity map of the six velocity components marked by colored rectangles in Figure \ref{fig:spec}. 
The solid orange contours show the VGPS 1.4 GHz radio continuum at 104--112 K levels in steps of 2 K from left (east) to right (west), while the dashed orange contours show the ATLASGAL 870 $\mu$m continuum (in steps of 0.3, 0.6 and 0.9 Jy $\rm beam^{-1}$). 
The red cross in panel (b) shows the position of ATLASGAL clump G49.111$-$0.322. 
}
\label{fig:hcop}
\end{figure*}

In total, we detected transitions from 24 molecular species, among which five (\hcop, \hcn, \cch, \occthreehtwo, and \hhco) show broadened emission line profiles in velocity range $+80$--$+110$ \kms, which is the LSR velocity range of the re-formed CO gas \citep{Koo_Interaction_1997}, while others are only detected around $\sim +61$ \kms\ which corresponds to the ambient MCs. 
The spectra of the five molecular transitions averaged in the entire mapped region, together with the \coa\ (hereafter CO if not specified) $J=1$--0 and 3--2 lines, are shown in Figure \ref{fig:spec}. 
The averaged spectra of other detected molecular species are shown in Appendix \ref{appendix}. 
The \hcop\ and \hcn\ lines exhibit the highest signal-to-noise ratio. 
Considering that the \hcn\ line may be affected by the hyperfine structures, we use the \hcop\ line to analyze the velocity structure in the mapped region. 

\par

We divided the entire \hcop\ spectrum into six components (see the colored rectangles in Figure \ref{fig:spec}) based on visual inspections on their spatial distributions. 
The integrated intensity map of the six components are shown in Figure \ref{fig:hcop}. 
Component (a) is bright toward the southwest of the mapped region and does not show spatial coincidence with the radio continuum. 
Component (b) is mainly located in the north and is spatially coincident with the 870 $\mu$m far-infrared continuum dust emission detected by the the APEX Telescope Large Area Survey of the GALaxy \citep[ATLASGAL,][]{Urquhart_ATLASGAL_2018}. 
The emission peak of \hcop\ in this component is also consistent with the ATLASGAL clump G49.111$-$0.322. 
Components (c) and (d) are contiguous in the spectrum: (c) is an emission peak while (d) exhibits a line wing structure (see Figure \ref{fig:spec}). 
However, component (c) is located toward the east of (d) and is spatially coincident with a bulge of radio continuum. 
Both components could be preshock gas of W51C or C-shocked gas. 
Components (e) and (f) are separated from the other velocity components and have been proposed to be the molecular gas re-formed behind fast J-shock. 
However, these two components show slightly different spatial distribution. 
The majority of the emission in component (e) is toward the center of the mapped region, while this component also contains a weak emission feature toward the west. 
For component (f), the emission toward the west vanishes, while the emission peak moves east compared with component (e). 
The spatial distribution of component (f) is similar to (c) and is also coincident with a bulge of radio continuum. 

\par

Although the re-formed gas can be divided into two velocity components, it is hard to perform spectral decomposition because the shocked gas does not necessarily exhibits Gaussian line profile and the signal-to-noise ratio of other transitions are limited. 
In the following analysis, we regard the two components as one. 
We also note that none of components (a) to (d) can be undoubtedly identified as the preshock gas. 
Besides, a pixel-by-pixel decomposition of all the observed spectra around 61 \kms\ is almost impossible. 
Therefore, we cannot estimate the properties of the preshock gas directly from our observation. 

\section{Discussion} \label{sec:dis}
\subsection{Estimation of the molecular column densities and abundance ratios} \label{sec:N}

\begin{figure}
\centering
\includegraphics[width=0.48\textwidth]{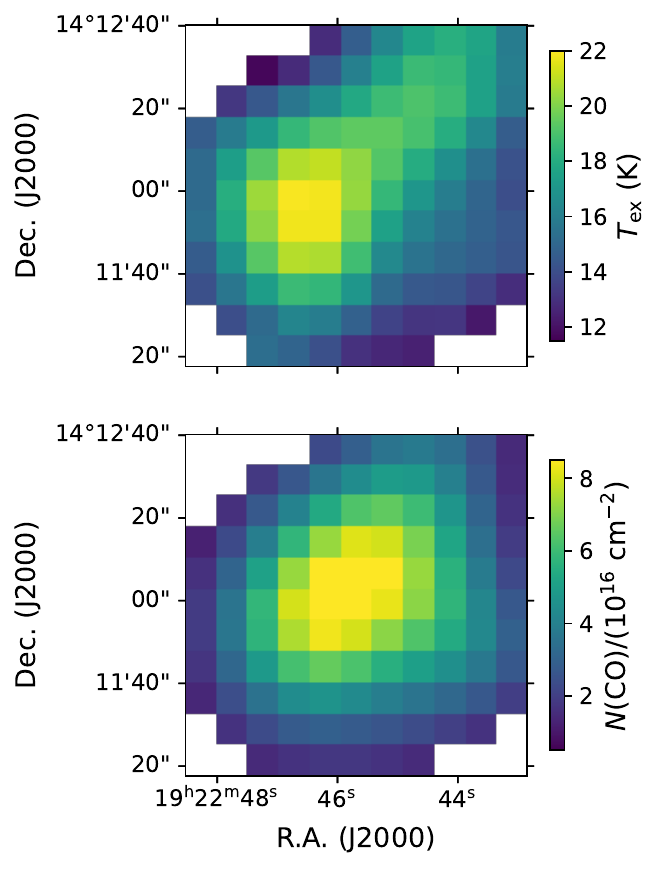}
\caption{Excitation temperature (upper panel) and column density (lower panel) maps of CO. 
}
\label{fig:Tex_N_CO}
\end{figure}

\begin{figure*}
\centering
\includegraphics[width=0.99\textwidth]{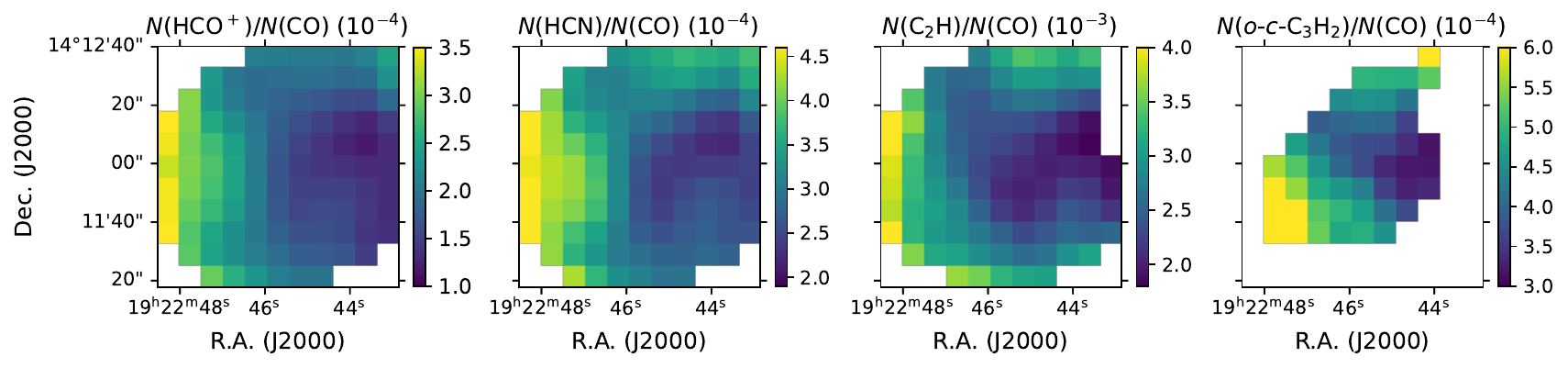}
\includegraphics[width=0.99\textwidth]{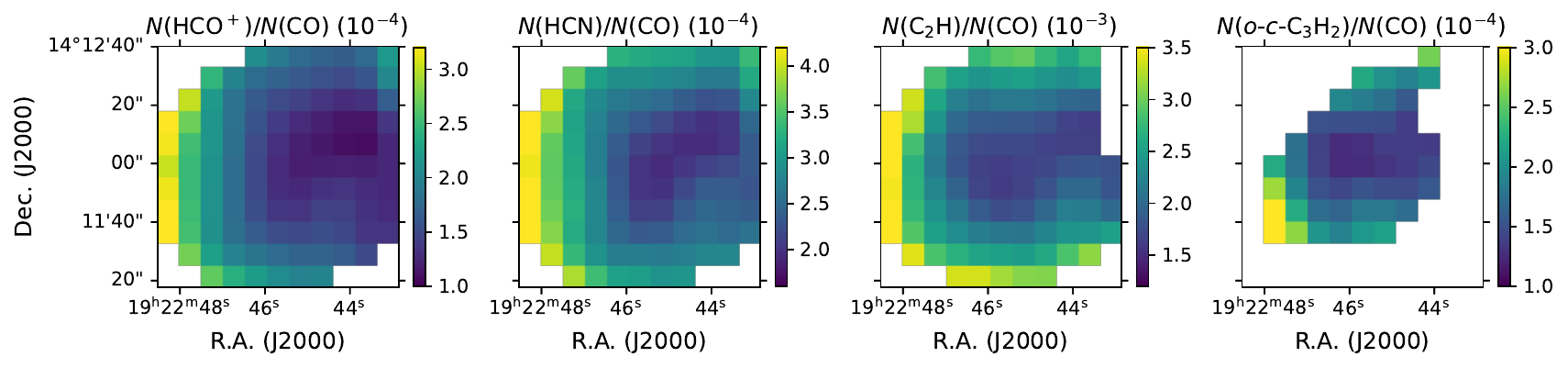}
\caption{Maps of abundance ratios \hcoponco, \hcnonco, \cchonco, and \octhtonco\ (from left to right) based on the assumptions $T_{\rm ex}=T_{\rm ex}({\rm CO})$ (upper row) and $T_{\rm ex}=5\rm \ K$ (lower row). 
}
\label{fig:N}
\end{figure*}

Considering that we do not have the data of multiple ($\geq3$) transitions of one molecular species, we assume local thermodynamic equilibrium (LTE) in our estimation of the molecular column densities. 
We assume that all the transitions in velocity range $+80$--$+110$ \kms\ are optically thin because of their large linewidths which in turn suggest large velocity gradients. 
The excitation temperature ($T_{\rm ex}$) of the CO lines can be estimated by fitting the relation \citep{Goldsmith_Population_1999}: 
\begin{equation}
    \log{\frac{N_u}{g_u}} = \log{N} - \log{Z} - \frac{E_u}{kT_{\rm ex}}
\end{equation}
where $N_u$ is the column density of the molecules in the upper energy level, $g_u$ is statistical weight of the upper energy level, $N$ is the total column density, $Z$ is the partition function, and $E_u$ is the energy of the upper level. 
In the optical thin limit, $N_u$ can be written as $N_u = (8\pi k\nu^2W)/(hc^3A_{ul})$ where $W$ is the integrated intensity and $A_{ul}$ is the Einstein $A$ coefficient of the transition. 
With $T_{\rm ex}$ obtained, the column density can be estimated via \citep{Mangum_How_2015}: 
\begin{equation} \label{eq:N}
    N = \frac{3k}{8\pi^3\nu} \frac{Z \exp{(E_{\rm u}/kT_{\rm ex})}}{S\mu^2} \frac{J_\nu(T_{\rm ex})}{J_\nu(T_{\rm ex})-J_\nu(T_{\rm bg})} W,
\end{equation}
where $S$ is the line strength, $\mu$ is the dipole moment, $T_{\rm bg}=2.73\ \rm K$ is the background temperature, and $J_\nu(T)$ is defined by $J_\nu(T)=(h\nu/k)/(\exp{(h\nu/kT)}-1)$. 
The molecular constants ($E_u$ and $S\mu^2$) are retrieved from Splatalogue\footnote{\url{https://splatalogue.online/}}. 
The obtained $T_{\rm ex}(\rm CO)$ and $N({\rm CO})$ maps are shown in Figure \ref{fig:Tex_N_CO}. 
Generally, the $T_{\rm ex}(\rm CO)$ is within the range 11.5--23 K, and the $N({\rm CO})$ is within $1\text{--}9\times 10^{16}\rm \ cm^{-2}$ throughout the clump. 
The peaks of both values are higher than those obtained by \citet{Koo_Interaction_1997a}, which are $T_{\rm ex}=11\rm \ K$ and $N({\rm CO})=4\times 10^{16}\rm \ cm^{-2}$, because the beam size of our data ($27^{\prime\prime}.6$) is smaller than that of their data ($55^{\prime\prime}$, which suffers more from the beam dilution effect because the angular size of the source is smaller than the beam size). 

\par

To estimate the column densities of other molecular species, we make two assumptions on their $T_{\rm ex}$: (1) assuming that $T_{\rm ex}$ of other species are equal to $T_{\rm ex}({\rm CO})$, and (2) fixing $T_{\rm ex}=5\rm \ K$ \citep{vanderTak_computer_2007} throughout the target region for other species because they trace denser gas compared with CO \citep{Shirley_Critical_2015}. 
We do not estimate the column density of $\rm H_2CO$ because of the low signal-to-noise ratio of its transition. 
The abundance ratio maps of \hcoponco, \hcnonco, \cchonco, and \octhtonco\ based on the two assumptions are separately shown in Figure \ref{fig:N}. 
Here we cannot obtain the abundances of \hcop, \hcn, \cch\ and \occthreehtwo\ because the abundance of CO is not necessarily $\sim 10^{-4}$ at such low column densities \citep[e.g.,][]{Burgh_Direct_2007}. 

\par

As can be seen in Figure \ref{fig:N}, the difference between the abundance ratios estimated for the two assumed cases of $T_{\rm ex}$ is within a factor of 2 for \hcop, \hcn\ and \cch, and a factor of 3 for \occthreehtwo. 
The obtained ranges of the abundance ratios, including the values obtained with the two assumptions, are 1.0--$4.0\times 10^{-4}$, 1.8--$5.3\times 10^{-4}$, 1.6--$5.0\times 10^{-3}$, and 1.2--$7.9\times 10^{-4}$ for \hcoponco, \hcnonco, \cchonco, and \octhtonco, respectively. 
The abundance ratios are all highest toward the eastern edge of the molecular clump and decrease toward the west, which is roughly consistent with the direction of the SNR blast wave. 

\par

We also estimate an upper limit of the \conco\ considering that no broadened [C\,I] emission (See Section \ref{sec:data2}) is detected. 
The upper limit of $N({\rm C^0})$ can be estimated with \citep{Izumi_Observations_2021}: 
\begin{equation}
\begin{aligned}
    N({\rm C^0}) =\ &4.67\times10^{16}\times \\
    &\frac{1+3\exp{(-23.6/T_{\rm ex})}+5\exp{(-62.5/T_{\rm ex})}}{1-\exp{(-23.6/T_{\rm ex})}} \int \tau_{\rm [C\, I]}\, dv
\end{aligned}
\end{equation}
where the optical depth of the [C\,I] line is: 
\begin{equation}
    \tau_{\rm [C\, I]} = -\ln{\left[ 1- \frac{T_{\rm mb}}{J_\nu(T_{\rm ex}) - J_\nu(T_{\rm bg})} \right] }. 
\end{equation}
We assume that the non-detected [C\,I] line profile follows a Gaussian with a full width at half maximum (FWHM) 15 \kms. 
We smooth the original data to a beam of $27^{\prime\prime}.6$ (see Section \ref{sec:data2}) and a velocity channel width of 3 \kms. 
The $3\sigma$ upper limit of the $T_{\rm mb}$ is 0.5 K. 
We vary the $T_{\rm ex}$ from 12 K to 22 K to find the maximum upper limit of $N(\rm C^0)$. 
Finally, we find an upper limit for \conco\ is $\lesssim 2$.

\subsection{Constraint on the gas density}

\begin{figure}
\centering
\includegraphics[width=0.35\textwidth]{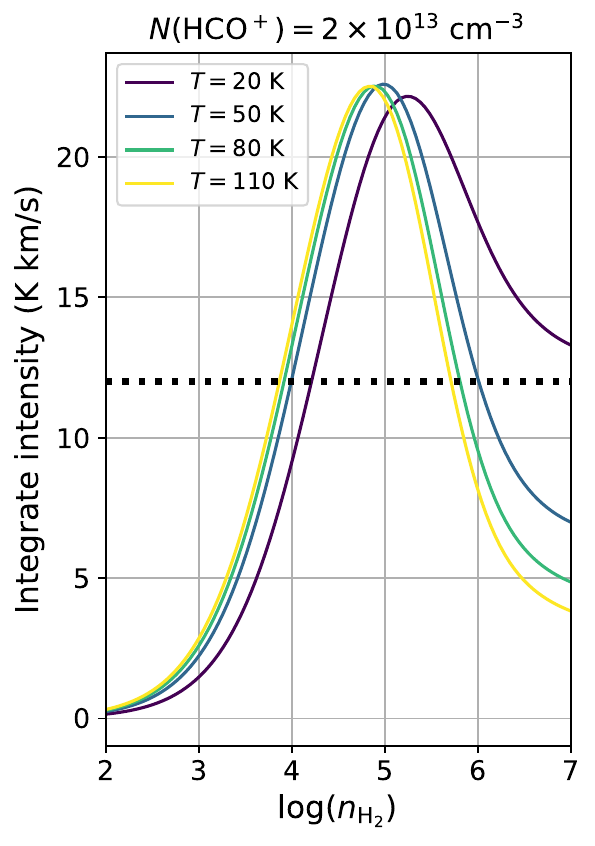}
\caption{The predicted integrated intensity of \hcop\ $J=1$--0 line as a function of $n_{\rm H_2}$. 
The lines with different colors show the models with different gas temperatures. 
The dotted black line in the right panel shows an integrated intensity of 12 K \kms. 
}
\label{fig:radex}
\end{figure}

Although we cannot conduct non-LTE analysis to fit the physical parameters of the molecular clump, we can still constrain its gas density. 
The peak integrated intensity of the \hcop\ $J=1$--0 line is $\approx 12\rm \ K\ km\ s^{-1}$ (which is inferred by combining Figure \ref{fig:hcop} (e) and (f)), and the peak $N(\rm HCO^+)$ in the LTE analysis is $\sim 2\times 10^{13}\rm \ cm^{-2}$ (which is inferred by combining Figure \ref{fig:Tex_N_CO} and \ref{fig:N}). 
We note that although the LTE analysis is indeed not precise, we consider $N(\rm HCO^+)$ for two cases of $T_{\rm ex}$ of 5 K and $\sim 20$ K as the lower and upper ends of the $T_{\rm ex}$ range separately. 
Therefore, the estimated peak can be regarded as a strict upper limit of the real $N(\rm HCO^+)$. 
To constrain the gas density, we run a set of physical parameters with the non-LTE radiative transfer code, RADEX \citep{vanderTak_computer_2007}, varying the gas density $n_{\rm H_2}$ and gas temperature $T$ with fixed $N(\rm HCO^+)=2\times 10^{13}\rm \ cm^{-2}$ and an FWHM of 15 \kms. 
Similar method has been used by \citet{Shirley_Critical_2015} to estimate the ``effective excitation density'' of dense gas tracers. 
We plot the predicted \sout{relation between} integrated intensity and the gas density at given values of $T$ in Figure \ref{fig:radex}. 
We find that the gas density should be at least $\sim 10^4\rm \ cm^{-3}$ to reproduce the observed peak intensity of the \hcop\ $J=1$--0 line. 
Although higher gas temperature also enhance the integrated intensity at $n_{\rm H_2}\lesssim 10^5\rm \ cm^{-3}$, this effect is not significant at $T\gtrsim50\rm \ K$. 

\subsection{Unusual \cchonco\ and \octhtonco\ abundance ratios}
We recall that the estimated abundance ratio is of order $\sim 10^{-3}$ for \cchonco\ and $\sim 10^{-4}$ for \octhtonco. 
In dense quiescent MCs, these ratios are typically $\sim 10^{-4}$--$10^{-5}$ and $\sim 10^{-5}$--$10^{-6}$, respectively \citep[e.g.,][]{Agundez_Chemistry_2013,Kim_ATLASGAL-selected_2020}. 
Therefore, the obtained \cchonco\ and \octhtonco\ ratios are higher than the typical values in dense MCs by 1--2 orders of magnitude. 

\par

Broadened emission line induced by SNR-MC interaction of \cch\ has only been found in SNRs IC443 \citep{vanDishoeck_Submillimeter_1993}, W28 \citep{Mazumdar_Submillimeter_2022a} and 3C391 \citep{Tu_Yebes_2024}, while broadened \ccthreehtwo\ line has only been found in SNR 3C391 \citep{Tu_Yebes_2024}. 
The \cchonco\ toward IC443 clump G, which is one of the prototypical molecular clumps interacting with SNRs, is $\sim 10^{-4}$ \citep{vanDishoeck_Submillimeter_1993}, similar to the value in typical quiescent MCs. 
In other environments of interstellar shocks, \cch\ and \ccthreehtwo\ are also seldom discussed. 
For example, these two species are regarded as tracers of ambient gas \citep[e.g.,][]{Shimajiri_Spectral-line_2015} or cavity walls exposed to UV radiation \citep[e.g.,][]{Tychoniec_Which_2021} in protostellar outflows. 
In molecular cloud core L1521, which is a source of shocked carbon chain chemistry with enhanced abundance of some carbon chain species, the abundance ratio \cthtonco\ is $\sim 10^{-5}$ \citep{Sato_New_1994,Liu_Search_2021}. 
We note that the ortho-to-para ratio of \ccthreehtwo\ is 3 in thermal equilibrium and is often observed to be within the range of 1--3 \citep{Park_Modeling_2006}. 
Therefore, the column density of \occthreehtwo\ should be similar to that of \ccthreehtwo, and the observed \cthtonco\ value toward L1251 is lower than the value we obtained in W51C by an order of magnitude. 

\par

Enhanced \cchonco\ ratio similar to our case was found in the circumnuclear disk of Seyfert galaxy NGC 1068 \citep{Viti_Molecular_2014,Nakajima_Molecular_2023}, but it was proposed to be the results of a complex interaction with shock and UV or X-ray radiation \citep{Garcia-Burillo_ALMA_2017}. 

\par 

Our observed abundance ratios are roughly consistent with the values in diffuse or translucent molecular gas: \citet{Liszt_CO_1998} for CO, \citet{Lucas_Plateau_1996} for \hcop, \citet{Liszt_Comparative_2001} for \hcn, and \citet{Lucas_Comparative_2000} for \cch\ and \ccthreehtwo. See also Table 3 of \citet{Liszt_Comparative_2006} for a compilation. We also refer to \citet{Kim_HyGAL_2023} for a later observation of all the species except CO. 
However, the gas density of the target clump ($\gtrsim 10^4\rm \ cm^{-3}$) is higher than the typical values of diffuse and translucent clouds \citep{Snow_Diffuse_2006}. 
Therefore, the dominating physical and chemical processes are expected to be different. 

\par

Carbon chain species, including \cch\ and \ccthreehtwo, are regarded as ``early-type species'' \citep{Sakai_Warm_2013,Taniguchi_Carbon-chain_2024}. 
The formation of carbon chain species rely strongly on $\rm C^+$ and $\rm C^0$ in the gas phase (see Figure 1 of \citet{Sakai_Warm_2013} and Figure 1 of \citet{Taniguchi_Carbon-chain_2024}), which are transformed to CO as the cloud evolves from diffuse gas to molecular cloud core. 
For instance, the formation routes of \cch\ include but are not limited to \citep{Taniguchi_Carbon-chain_2024}: 
\begin{equation} \label{react:C2Hroutes}
\begin{aligned}
    & (a)\ \rm C^+ \xrightarrow{H_2} CH_2^+ \xrightarrow{H_2} CH_3^+ \xrightarrow{e^-} CH_2 \xrightarrow{C^+} C_2H \\
    & (b)\ \rm C^+ \xrightarrow{H_2} CH_2^+ \xrightarrow{H_2} CH_3^+ \xrightarrow{C} C_2H_2^+ \xrightarrow{e^-} C_2H \\
    & (c)\ \rm C^+ \xrightarrow{H_2} CH_2^+ \xrightarrow{e^-} CH \xrightarrow{C^+} C_2^+ \xrightarrow{H_2} C_2H^+ \xrightarrow{H_2} C_2H_2^+ \xrightarrow{e^-} C_2H. 
\end{aligned}
\end{equation}
Therefore, carbon chains are formed efficiently at the earliest phase of MC formation and are depleted or destructed in evolved MCs, which has been verified by many chemical simulations \citep[e.g.,][]{Suzuki_Survey_1992,Taniguchi_Investigation_2019}. 
Since the target clump is molecular gas re-formed behind a dissociative J-shock, it is supposed to be at the earliest phase of MC evolution with abundant $\rm C^+$ and $\rm C^0$ in the gas phase.
Therefore, enhanced abundances of carbon chain species, and in turn their abundance ratios to CO, are possible. 
Simulation of \citet{Neufeld_Fast_1989} also predict a plateau of \cch\ abundance $X(\rm C_2H)$ soon after the J-shock. 
However, this simulation only considered fast J-shock in dense preshock gas $> 10^4\rm \ cm^{-3}$ which is higher than the preshock density in our case, and did not either predict significantly enhanced \cchonco\ abundance ratio. 

\subsection{Chemical simulation of molecular re-formation behind J-shock}
To further investigate whether the enhanced \cchonco\ and \octhtonco\ abundance ratios are due to the chemistry induced by J-shock, we use the Paris-Durham shock code \citep{Flower_Theoretical_1985,Flower_influence_2003,Flower_Interpreting_2015,Godard_Models_2019} to simulate the molecular re-formation behind J-shock. 
The code is a public numerical tool to compute the coupled dynamical, thermal, and chemical evolution of interstellar medium subject to plane parallel shock wave. 
The computation consists of two steps. 
In step 1, the code simulates the evolution of a quiescent cloud and evolve into a steady state. 
In step 2, the code uses the outputs of step 1 as the preshock conditions (including both physical and chemical parameters) and allows a shock wave to propagate into the preshock cloud. 

\par

We simulate both irradiated and non-irradiated shocks. 
In the irradiated case, we assume that a cloud with a visual extinction of $A_{\rm V}=2$ (at the interface between translucent and dense MC \citep{Snow_Diffuse_2006}) is irradiated by the interstellar radiation field with $G_0=1.6$ in the Habing unit\footnote{$G_0=1$ indicates an integrated flux of $1.6\times 10^{-3} \rm \ erg\ cm^{-2}\ s^{-1}$ in 6--13.6 eV.} \citep{Parravano_Time_2003,Wolfire_Photodissociation_2022}. 
The magnetic field in the simulation is controlled by a parameter $\beta = B(\mu{\rm G})/\sqrt{n_{\rm H}(\rm cm^{-3})}$. 
We adopt $\beta=1$ which is close to the magnetic field in interstellar clouds \citep{Crutcher_Magnetic_2010}. 
The cosmic-ray ionization rate per $\rm H_2$ is fixed to be $1.3\times 10^{-17}\rm \ s^{-1}$ in step 1, which is the typical value in MCs \citep{Caselli_Ionization_1998}, and is enhanced to $5\times 10^{-16} \rm \ s^{-1}$ in step 2, roughly consistent with previous observations and simulations \citep{Ceccarelli_Supernova-enhanced_2011,Shingledecker_Inference_2016,Yamagishi_Cosmic-ray-driven_2023}. 
The propagation time of the shock is limited to $t<3\times 10^4\rm \ yr$, which is consistent with previous constraint on the age of W51C \citep{Koo_ROSAT_1995,Park_Shells_2013}. 
We vary the preshock H nucleus density $n_{\rm H}$ to be $2\times 10^2$, $2\times 10^{3}$ and $2\times 10^4\rm \ cm^{-3}$, and shock velocity $V_{\rm s}$ to be 25, 30, 40, 50 and 60 \kms. 
At lower $V_{\rm s}$, the shock becomes C-shock. 
We find that the preshock density $n_{\rm H}=2\times 10^3\rm \ cm^{-3}$ can better reproduce the observed abundance ratio and the results that the target clump has a density $n_{\rm H_2} \gtrsim 10^4\rm \ cm^{-3}$. 
Therefore we limit our discussion with $n_{\rm H}=2\times 10^3\rm \ cm^{-3}$. 

\begin{figure*}
\centering
\includegraphics[width=0.95\textwidth]{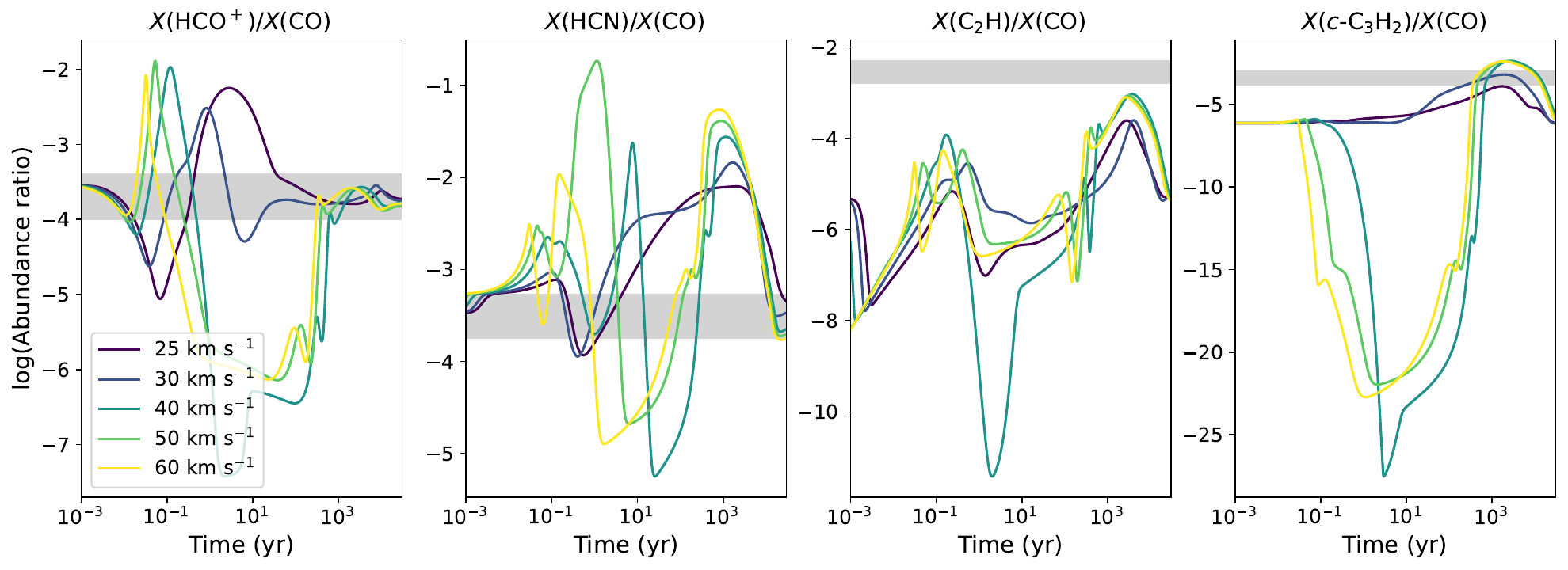}
\includegraphics[width=0.95\textwidth]{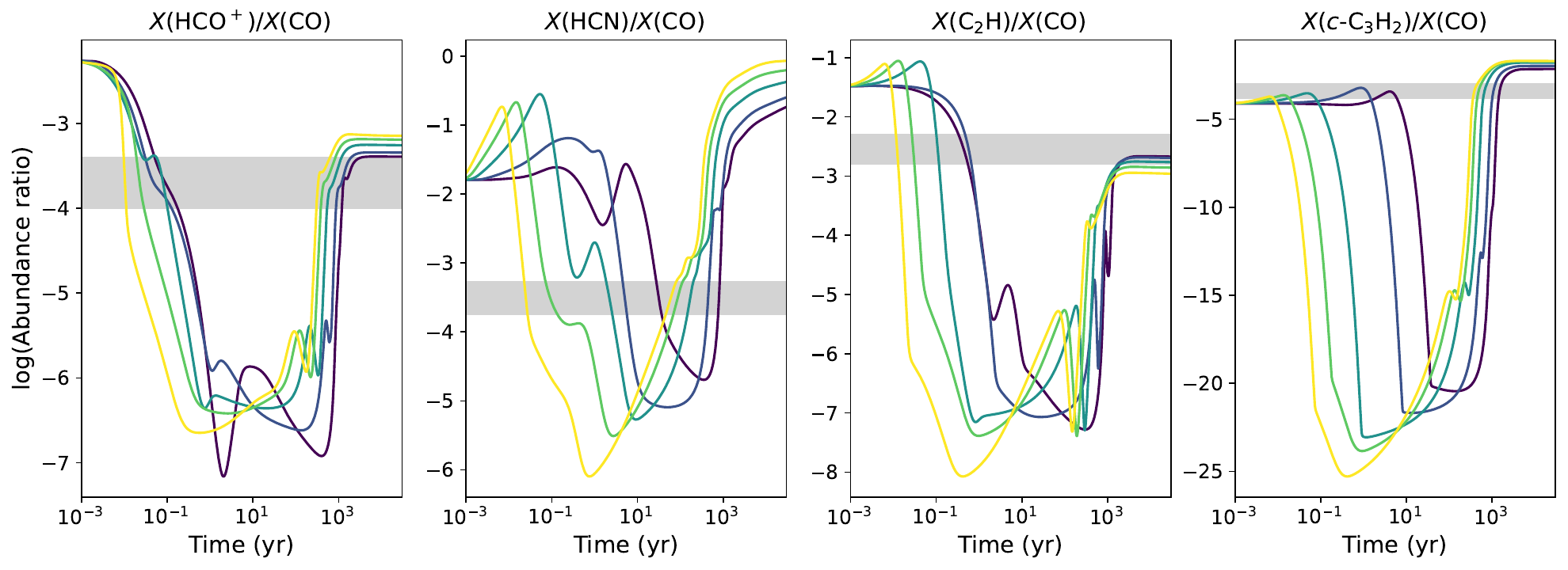}
\caption{Abundance ratios \hcoponco, \hcnonco, \cchonco\ and \cthtonco\ (from left to right) predicted by the Paris-Durham shock code as a function of time with a preshock density of $n_{\rm H}=2\times 10^3\rm \ cm^{-3}$. 
The upper row shows the results of non-irradiated shock, while the lower row shows the results of shock irradiated by an interstellar radiation field ($G_0=1.6$) with a visual extinction of $A_{\rm V}=2$. 
Lines with different colors shows the results of different shock velocities as shown in the labels of the upper right panel. 
The gray shaded regions delineate the range obtained from our observation. 
}
\label{fig:model}
\end{figure*}

\begin{figure*}
\centering
\includegraphics[height=0.3\textheight]{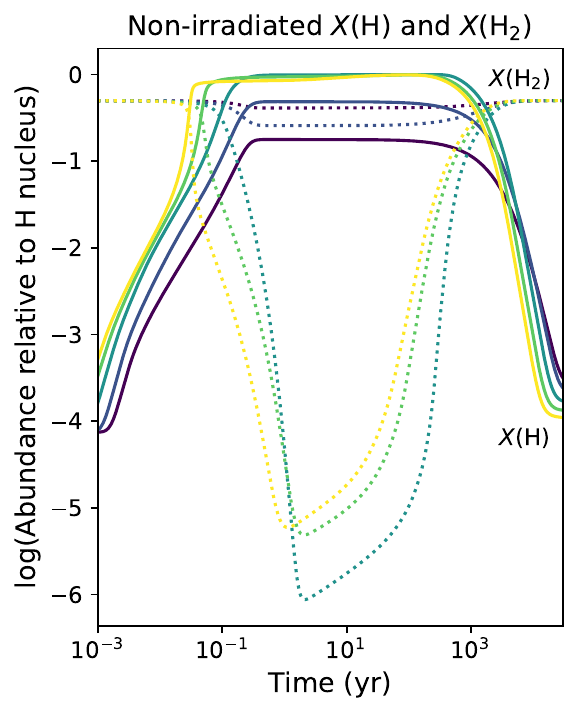}
\includegraphics[height=0.3\textheight]{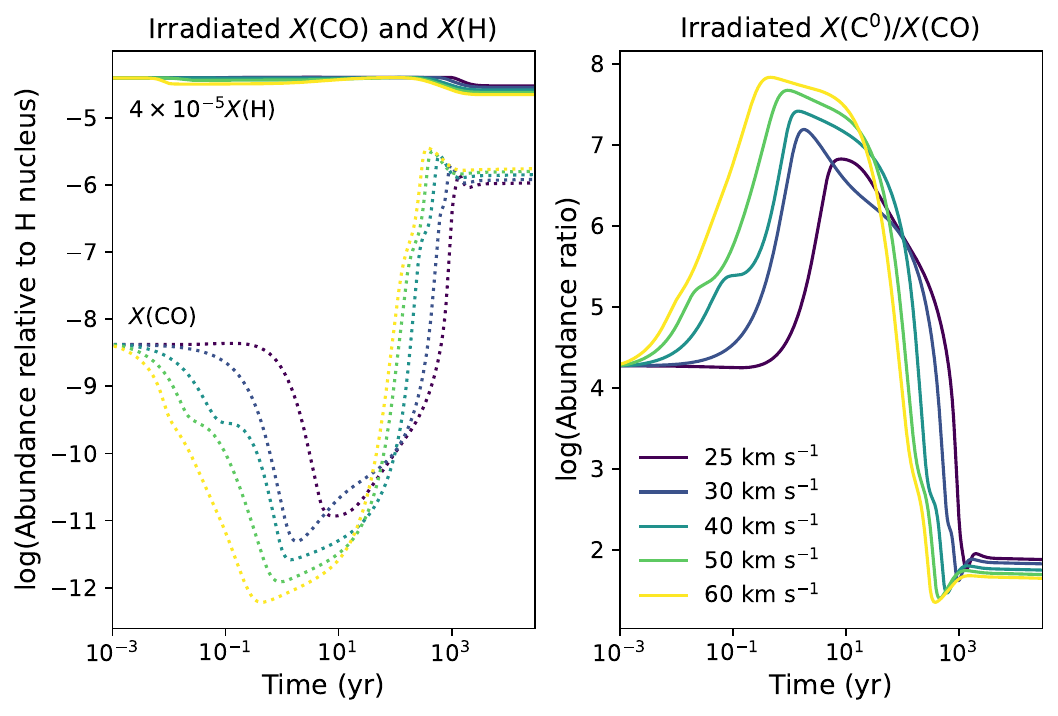}
\caption{Auxiliary results of the non-irradiated (left panel) and irradiated (middle and right panel) shock models with a preshock density of $n_{\rm H}=2\times 10^3\rm \ cm^{-3}$. 
Lines with different colors shows the results of different shock velocity as shown in the labels of the right panel. 
\textit{left panel:} Evolution of the abundances (denoted by $X$) of H (solid lines) and $\rm H_2$ (dotted lines) relative to the total H nucleus in the non-irradiated models. 
\textit{middle panel:} Evolution of the abundances of H (solid lines, multiplied by a factor of $4\times 10^{-5}$) and CO (dotted lines) relative to the total H nucleus in the irradiated models. 
\textit{right panel:} Evolution of the abundance ratio \conco\ in the irradiated models. 
}
\label{fig:model_problem}
\end{figure*}

Figure \ref{fig:model} shows the abundance ratios predicted by the non-irradiated (upper row) and irradiated (lower row) shock models. 
For the non-irradiated shock model, all of the four abundance ratios except \cchonco\ can be well reproduced, while the difference between the peak values of the modeled \cchonco\ and observed range is within an order of magnitude. 
Therefore, we conclude that the observed abundance ratio can essentially be reproduced by the J-shock chemistry. 
We note that the \cchonco\ and \cthtonco\ reach the peak values at $\sim 3\times 10^3$ yr, and then quickly decrease. 
This is consistent with the scenario that the carbon chain species are ``early-type species'' and are destructed along with the evolution of the MC \citep{Sakai_Warm_2013,Taniguchi_Carbon-chain_2024}. 
Specifically, \cch\ is mainly formed via: 
\begin{equation}
    \rm CH^+ \xrightarrow{H_2} CH_2^+ \xrightarrow{H_2} CH_3^+ \xrightarrow{C} C_2H_2^+ \xrightarrow{e^-} C_2H, 
\end{equation}
at $t \lesssim 3$ kyr, where the $\rm CH^+$ can be formed through either $\rm C^++H_2$ or $\rm C+H_3^+$. 
This pathway is similar to route $(b)$ mentioned in reactions \ref{react:C2Hroutes}. 
Atomic C, which is key to the formation of $\rm C_2H_2^+$, is formed in the J-shock through the dissociation of CO and CH. 
At $t \gtrsim 3$ kyr, \cch\ is mainly destructed by atomic O to form CO. 
The formation of \ccthreehtwo\ is much more complicated, ended with $\rm \textit{c-}C_3H_3^+ + e^- \rightarrow \textit{c-}C_3H_2 + H$, where the $\rm \textit{c-}C_3H_3^+$ molecule can be formed via $\textit{c-}{\rm C_3H^+}/\textit{c-}{\rm C_3H_2^+} + \rm H_2$. 

\par

For the irradiated shock model, the \hcoponco\ and \cchonco\ can be well reproduced, while the \cthtonco\ is within an order of magnitude away from the observed range. 
However, the \hcnonco\ is strongly overestimated by the irradiated model. 
Therefore, although this model can roughly reproduce the observed abundance ratios, its performance is worse than the non-irradiated shock model. 
The formation pathway of \cch\ in the irradiated shock model is slightly different from that in the non-irradiated shock model: 
\begin{equation}
    \rm C^+ \xrightarrow{H_2} CH^+ \xrightarrow{C} C_2^+ \xrightarrow{H_2} C_2H^+ \xrightarrow{H_2} C_2H_2^+ \xrightarrow{e^-} C_2H, 
\end{equation}
because of the higher abundance of C and $\rm C^+$ due to the external UV field. 

\par

We note that although both models can reproduce the observed abundance ratios to some extent, the simulation is not exempt from problems. 
We recall that there are some other constraints on the physical properties of the target molecular clump. 
\citet{Koo_Interaction_1997} found that the abundance ratio $N({\rm CO})/N({\rm H\, I})$ is $<4\times 10^{-5}$, which is the expected value if CO has totally been re-formed while HI has yet to form $\rm H_2$. 
To explain this effect, they proposed that the hydrogen nuclei are mainly in the atomic form instead of molecular form according to the $N(\rm H\, I)$ vs. $N(\rm CO)$ scatter plot (see their Figure 5). 
However, as plotted in the left panel of Figure \ref{fig:model_problem}, a significant fraction of H has recombined to $\rm H_2$ even before $2\times 10^3$ yr regardless of shock velocity in the non-irradiated shock models. 
Therefore, the non-irradiated shock models can hardly reproduce the trends proposed by \citet{Koo_Interaction_1997}. 
On the other hand, for the irradiated models, the $N({\rm CO})/N({\rm H\, I}) <4\times 10^{-5}$ can be naturally explained by the fact that neither CO nor $\rm H_2$ has been remarkably re-formed (see the middle panel of Figure \ref{fig:model_problem}). 
But in Section \ref{sec:N} we obtain an upper limit for the \conco\ abundance ratio $\lesssim 2$. 
However, the irradiated shock models may have significantly overestimated the \conco\ abundance ratio because of the photodissociation of incident UV radiation. 
This could be because the target clump suffers from a less intense UV field compared with our assumption. 
An alternative explanation is the uncertainty in the treatment of UV radiative transfer in the shock model as shown in \citet{Godard_Models_2019}. 

\par

In summary, our illustrative simulation can essentially reproduce, despite some defects, the observed abundance ratios, especially the enhanced \cchonco\ and \cthtonco. 
This suggests that these abundance ratios can be explained by the chemistry induced by the J-shock. 
However, we also note that the chemical simulation is not exempt from problems such as the lack of consideration of three-dimensional structure of the MC and incomplete chemical network. 
In addition, we only investigate some specific cases with simple assumptions on the properties such as UV field, magnetic field, and CR ionization rate, instead of exploring the full parameter space.
Further improvement on the chemical network, especially the grain-surface chemistry that strongly affects the initial condition of the shock and plays an important role in carbon-chain chemistry \citep{Sakai_Warm_2013}, and the treatment of UV radiative transfer may be helpful for the shock model to better reproduce the observation, but this is beyond the scope of this paper.

\section{Conclusions} \label{sec:con}
We present new observation toward W51C clump 2 in 71.4--89.7 GHz with the Yebes 40 m radio telescope to study the molecular chemistry induced by J-shock. 
Five molecular species (\hcop, \hcn, \cch, \occthreehtwo, and $\rm H_2CO$) exhibit broadened emission line profiles in $+80$--$+110$ \kms\ which is the velocity range of the re-formed molecular gas behind J-shock. 
We find that the spectrum of \hcop\ $J=1$--0 can be divided into six velocity components, in which the re-formed gas can be divided into two components with different spatial distribution. 
To facilitate the analysis, we regard the two re-formed gas components as one. 

\par

With the CO $J=1$--0 and 3--2 data, we estimate the excitation temperature and column density of CO based on the LTE assumption, and obtain the abundance ratios of \hcop, \hcn, \cch, \occthreehtwo\ to CO, which are $N({\rm HCO^+})/N({\rm CO})\sim (1.0\text{--}4.0)\times 10^{-4}$, $N({\rm HCN})/N({\rm CO})\sim (1.8\text{--}5.3)\times 10^{-4}$, $N({\rm C_2H})/N({\rm CO})\sim (1.6\text{--}5.0)\times 10^{-3}$, and $N({o\text{-}c\text{-}{\rm C_3H_2}})/N({\rm CO})\sim (1.2\text{--}7.9)\times 10^{-4}$. 
We also obtain an upper limit for the \conco\ abundance ratio $\lesssim 2$. 
We find that the \cchonco\ and \occthreehtwo\ ratios are higher than the typical values in dense MCs and the value in SNR IC443 clump G. 
This enhancement can be qualitatively attributed to the chemistry at the earliest phase of MC formation when abundant $\rm C^+$ and C in the gas phase boost the formation of carbon chain species. 

\par

To further investigate whether the enhanced \cchonco\ and \cthtonco\ are due to J-shock chemistry, we conduct a chemical simulation with the Paris-Durham shock code. 
We find that both the non-irradiated and the irradiated shock models with a preshock density of $n_{\rm H}=2\times 10^3 \rm \ cm^{-3}$ can basically reproduce the observed abundance ratios despite some defects, which suggests that the abundance ratios can indeed be attributed to the chemistry induced by J-shock. 
However, the non-irradiated models overestimate the re-formation of $\rm H_2$, while the irradiated models overestimate the \conco\ abundance ratio. 
Improvements on the shock code may be helpful for reproducing the observation.

\begin{acknowledgements}
The authors thank Alba Vidal García who carried out the observations and the first inspection of the data quality, other staff of the Yebes observatory who supported the observation and data transmission, Mitsuyoshi Yamagishi who provided the data cubes of \coa\ and [C\,I], and Benjamin Godard who offered helps on the use of the Paris-Durham shock code. 
V.W. acknowledges the CNRS program “Physique et Chimie du Milieu Interstellaire” (PCMI) co-funded by the Centre National d’Etudes Spatiales (CNES).
Y.C. acknowledges the support from NSFC grants Nos. 12173018 and 12121003. 
P.Z. acknowledges the support from NSFC grant No. 12273010. 
This article is based on observations carried out with the Yebes 40 m telescope (project code: 24A003). The 40 m radio telescope at Yebes Observatory is operated by the Spanish Geographic Institute (IGN; Ministerio de Transportes y Movilidad Sostenible). 
This work has made use of the Paris-Durham public shock code V1.1, distributed by the CNRS-INSU National Service “ISM Platform” at the Paris Observatory Data Center (\url{http://ism.obspm.fr}). 
\end{acknowledgements}

%
%

\bibliography{article_W51C_Yebes}

\begin{thebibliography}{76}
\expandafter\ifx\csname natexlab\endcsname\relax\def\natexlab#1{#1}\fi

\bibitem[{Ag{\'u}ndez \& Wakelam(2013)}]{Agundez_Chemistry_2013}
Ag{\'u}ndez, M. \& Wakelam, V. 2013, ChRv, 113, 8710

\bibitem[{{Astropy Collaboration} {et~al.}(2022){Astropy Collaboration},
  {Price-Whelan}, Lim, Earl, Starkman, Bradley, Shupe, Patil, Corrales,
  Brasseur, N{\"o}the, Donath, Tollerud, Morris, Ginsburg, Vaher, Weaver,
  Tocknell, Jamieson, {van Kerkwijk}, Robitaille, Merry, Bachetti, G{\"u}nther,
  Aldcroft, {Alvarado-Montes}, Archibald, B{\'o}di, Bapat, Barentsen,
  Baz{\'a}n, Biswas, Boquien, Burke, Cara, Cara, Conroy, Conseil, Craig, Cross,
  Cruz, D'Eugenio, Dencheva, Devillepoix, Dietrich, Eigenbrot, Erben, Ferreira,
  {Foreman-Mackey}, Fox, Freij, Garg, Geda, Glattly, Gondhalekar, Gordon,
  Grant, Greenfield, Groener, Guest, Gurovich, Handberg, Hart,
  {Hatfield-Dodds}, Homeier, Hosseinzadeh, Jenness, Jones, Joseph, Kalmbach,
  Karamehmetoglu, Ka{\l}uszy{\'n}ski, Kelley, Kern, Kerzendorf, Koch, Kulumani,
  Lee, Ly, Ma, MacBride, Maljaars, Muna, Murphy, Norman, O'Steen, Oman,
  Pacifici, Pascual, {Pascual-Granado}, Patil, Perren, Pickering, Rastogi,
  Roulston, Ryan, Rykoff, Sabater, Sakurikar, Salgado, Sanghi, Saunders,
  Savchenko, Schwardt, {Seifert-Eckert}, Shih, Jain, Shukla, Sick, Simpson,
  Singanamalla, Singer, Singhal, Sinha, Sip{\H o}cz, Spitler, Stansby,
  Streicher, {\v S}umak, Swinbank, Taranu, Tewary, Tremblay, {de Val-Borro},
  Van~Kooten, Vasovi{\'c}, Verma, {de Miranda Cardoso}, Williams, Wilson,
  Winkel, {Wood-Vasey}, Xue, Yoachim, Zhang, Zonca, \& {Astropy Project
  Contributors}}]{AstropyCollaboration_Astropy_2022}
{Astropy Collaboration}, {Price-Whelan}, A.~M., Lim, P.~L., {et~al.} 2022, ApJ,
  935, 167

\bibitem[{{Astropy Collaboration} {et~al.}(2018){Astropy Collaboration},
  {Price-Whelan}, Sip{\H o}cz, G{\"u}nther, Lim, Crawford, Conseil, Shupe,
  Craig, Dencheva, Ginsburg, VanderPlas, Bradley, {P{\'e}rez-Su{\'a}rez}, {de
  Val-Borro}, Aldcroft, Cruz, Robitaille, Tollerud, Ardelean, Babej, Bach,
  Bachetti, Bakanov, Bamford, Barentsen, Barmby, Baumbach, Berry, Biscani,
  Boquien, Bostroem, Bouma, Brammer, Bray, Breytenbach, Buddelmeijer, Burke,
  Calderone, Cano~Rodr{\'i}guez, Cara, Cardoso, Cheedella, Copin, Corrales,
  Crichton, D'Avella, Deil, Depagne, Dietrich, Donath, Droettboom, Earl, Erben,
  Fabbro, Ferreira, Finethy, Fox, Garrison, Gibbons, Goldstein, Gommers, Greco,
  Greenfield, Groener, Grollier, Hagen, Hirst, Homeier, Horton, Hosseinzadeh,
  Hu, Hunkeler, Ivezi{\'c}, Jain, Jenness, Kanarek, Kendrew, Kern, Kerzendorf,
  Khvalko, King, Kirkby, Kulkarni, Kumar, Lee, Lenz, Littlefair, Ma, Macleod,
  Mastropietro, McCully, Montagnac, Morris, Mueller, Mumford, Muna, Murphy,
  Nelson, Nguyen, Ninan, N{\"o}the, Ogaz, Oh, Parejko, Parley, Pascual, Patil,
  Patil, Plunkett, Prochaska, Rastogi, Reddy~Janga, Sabater, Sakurikar,
  Seifert, Sherbert, {Sherwood-Taylor}, Shih, Sick, Silbiger, Singanamalla,
  Singer, Sladen, Sooley, Sornarajah, Streicher, Teuben, Thomas, Tremblay,
  Turner, Terr{\'o}n, {van Kerkwijk}, {de la Vega}, Watkins, Weaver, Whitmore,
  Woillez, Zabalza, \& {Astropy
  Contributors}}]{AstropyCollaboration_Astropy_2018}
{Astropy Collaboration}, {Price-Whelan}, A.~M., Sip{\H o}cz, B.~M., {et~al.}
  2018, ApJ, 156, 123

\bibitem[{Bachiller {et~al.}(2001)Bachiller, P{\'e}rez~Guti{\'e}rrez, Kumar, \&
  Tafalla}]{Bachiller_Chemically_2001}
Bachiller, R., P{\'e}rez~Guti{\'e}rrez, M., Kumar, M. S.~N., \& Tafalla, M.
  2001, A\&A, 372, 899

\bibitem[{Benedettini {et~al.}(2012)Benedettini, Busquet, Lefloch, Codella,
  Cabrit, Ceccarelli, Giannini, Nisini, Vasta, Cernicharo, Lorenzani, \& {di
  Giorgio}}]{Benedettini_CHESS_2012}
Benedettini, M., Busquet, G., Lefloch, B., {et~al.} 2012, A\&A, 539, L3

\bibitem[{Brogan {et~al.}(2013)Brogan, Goss, Hunter, Richards, Chandler,
  Lazendic, Koo, Hoffman, \& Claussen}]{Brogan_OH_2013}
Brogan, C.~L., Goss, W.~M., Hunter, T.~R., {et~al.} 2013, ApJ, 771, 91

\bibitem[{Burgh {et~al.}(2007)Burgh, France, \& McCandliss}]{Burgh_Direct_2007}
Burgh, E.~B., France, K., \& McCandliss, S.~R. 2007, ApJ, 658, 446

\bibitem[{Burkhardt {et~al.}(2019)Burkhardt, Shingledecker, Le~Gal, McGuire,
  Remijan, \& Herbst}]{Burkhardt_Modeling_2019}
Burkhardt, A.~M., Shingledecker, C.~N., Le~Gal, R., {et~al.} 2019, ApJ, 881, 32

\bibitem[{Caselli {et~al.}(1998)Caselli, Walmsley, Terzieva, \&
  Herbst}]{Caselli_Ionization_1998}
Caselli, P., Walmsley, C.~M., Terzieva, R., \& Herbst, E. 1998, ApJ, 499, 234

\bibitem[{Ceccarelli {et~al.}(2011)Ceccarelli, {Hily-Blant}, Montmerle, Dubus,
  Gallant, \& Fiasson}]{Ceccarelli_Supernova-enhanced_2011}
Ceccarelli, C., {Hily-Blant}, P., Montmerle, T., {et~al.} 2011, ApJ, 740, L4

\bibitem[{Codella {et~al.}(2020)Codella, Ceccarelli, Bianchi, Balucani, Podio,
  Caselli, Feng, Lefloch, {L{\'o}pez-Sepulcre}, Neri, Spezzano, \&
  De~Simone}]{Codella_Seeds_2020}
Codella, C., Ceccarelli, C., Bianchi, E., {et~al.} 2020, A\&A, 635, A17

\bibitem[{Crutcher {et~al.}(2010)Crutcher, Wandelt, Heiles, Falgarone, \&
  Troland}]{Crutcher_Magnetic_2010}
Crutcher, R.~M., Wandelt, B., Heiles, C., Falgarone, E., \& Troland, T.~H.
  2010, ApJ, 725, 466

\bibitem[{Cuppen {et~al.}(2010)Cuppen, Kristensen, \& Gavardi}]{Cuppen_H2_2010}
Cuppen, H.~M., Kristensen, L.~E., \& Gavardi, E. 2010, MNRAS, 406, L11

\bibitem[{Draine \& McKee(1993)}]{Draine_Theory_1993}
Draine, B.~T. \& McKee, C.~F. 1993, ARA\&A, 31, 373

\bibitem[{Dumas {et~al.}(2014)Dumas, Vaupr{\'e}, Ceccarelli, {Hily-Blant},
  Dubus, Montmerle, \& Gabici}]{Dumas_Localized_2014}
Dumas, G., Vaupr{\'e}, S., Ceccarelli, C., {et~al.} 2014, ApJ, 786, L24

\bibitem[{Flower \& {Pineau des For{\^e}ts}(2003)}]{Flower_influence_2003}
Flower, D.~R. \& {Pineau des For{\^e}ts}, G. 2003, MNRAS, 343, 390

\bibitem[{Flower \& {Pineau des For{\^e}ts}(2015)}]{Flower_Interpreting_2015}
Flower, D.~R. \& {Pineau des For{\^e}ts}, G. 2015, A\&A, 578, A63

\bibitem[{Flower {et~al.}(1985)Flower, {Pineau des For{\^e}ts}, \&
  Hartquist}]{Flower_Theoretical_1985}
Flower, D.~R., {Pineau des For{\^e}ts}, G., \& Hartquist, T.~W. 1985, MNRAS,
  216, 775

\bibitem[{{Garc{\'i}a-Burillo} {et~al.}(2017){Garc{\'i}a-Burillo}, Viti,
  Combes, Fuente, Usero, Hunt, Mart{\'i}n, Krips, Aalto, Aladro, Ramos~Almeida,
  {Alonso-Herrero}, Casasola, Henkel, Querejeta, Neri, Costagliola, Tacconi, \&
  {van der Werf}}]{Garcia-Burillo_ALMA_2017}
{Garc{\'i}a-Burillo}, S., Viti, S., Combes, F., {et~al.} 2017, A\&A, 608, A56

\bibitem[{Godard {et~al.}(2019)Godard, {Pineau des For{\^e}ts}, Lesaffre,
  Lehmann, Gusdorf, \& Falgarone}]{Godard_Models_2019}
Godard, B., {Pineau des For{\^e}ts}, G., Lesaffre, P., {et~al.} 2019, A\&A,
  622, A100

\bibitem[{Goldsmith \& Langer(1999)}]{Goldsmith_Population_1999}
Goldsmith, P.~F. \& Langer, W.~D. 1999, ApJ, 517, 209

\bibitem[{{G{\'o}mez-Ruiz} {et~al.}(2015){G{\'o}mez-Ruiz}, Codella, Lefloch,
  Benedettini, Busquet, Ceccarelli, Nisini, Podio, \&
  Viti}]{Gomez-Ruiz_density_2015}
{G{\'o}mez-Ruiz}, A.~I., Codella, C., Lefloch, B., {et~al.} 2015, MNRAS, 446,
  3346

\bibitem[{Green {et~al.}(1997)Green, Frail, Goss, \&
  Otrupcek}]{Green_Continuation_1997}
Green, A.~J., Frail, D.~A., Goss, W.~M., \& Otrupcek, R. 1997, ApJ, 114, 2058

\bibitem[{Green(2019)}]{Green_revised_2019}
Green, D.~A. 2019, JApA, 40, 36

\bibitem[{Gusdorf {et~al.}(2008)Gusdorf, Cabrit, Flower, \& Pineau
  Des~For{\^e}ts}]{Gusdorf_SiO_2008}
Gusdorf, A., Cabrit, S., Flower, D.~R., \& Pineau Des~For{\^e}ts, G. 2008,
  A\&A, 482, 809

\bibitem[{Hollenbach {et~al.}(2013)Hollenbach, Elitzur, \&
  McKee}]{Hollenbach_Interstellar_2013}
Hollenbach, D., Elitzur, M., \& McKee, C.~F. 2013, ApJ, 773, 70

\bibitem[{Hollenbach \& McKee(1989)}]{Hollenbach_Molecule_1989}
Hollenbach, D. \& McKee, C.~F. 1989, ApJ, 342, 306

\bibitem[{Izumi {et~al.}(2021)Izumi, Fukui, Tachihara, Fujita, Torii, Kamazaki,
  Kaneko, Silva, Iono, Momose, Sugimoto, Nakazato, Kosugi, Maekawa, Takahashi,
  Yoshino, \& Asayama}]{Izumi_Observations_2021}
Izumi, N., Fukui, Y., Tachihara, K., {et~al.} 2021, PASJ, 73, 174

\bibitem[{Kim {et~al.}(2023)Kim, Schilke, Neufeld, Jacob, {S{\'a}nchez-Monge},
  Seifried, Godard, Menten, Walch, Falgarone, Veena, Bialy, M{\"o}ller, \&
  Wyrowski}]{Kim_HyGAL_2023}
Kim, W.~J., Schilke, P., Neufeld, D.~A., {et~al.} 2023, A\&A, 670, A111

\bibitem[{Kim {et~al.}(2020)Kim, Wyrowski, Urquhart, {P{\'e}rez-Beaupuits},
  Pillai, Tiwari, \& Menten}]{Kim_ATLASGAL-selected_2020}
Kim, W.~J., Wyrowski, F., Urquhart, J.~S., {et~al.} 2020, A\&A, 644, A160

\bibitem[{Koo {et~al.}(1995)Koo, Kim, \& Seward}]{Koo_ROSAT_1995}
Koo, B.-C., Kim, K.-T., \& Seward, F.~D. 1995, ApJ, 447, 211

\bibitem[{Koo \& Moon(1997{\natexlab{a}})}]{Koo_Interaction_1997a}
Koo, B.-C. \& Moon, D.-S. 1997{\natexlab{a}}, ApJ, 475, 194

\bibitem[{Koo \& Moon(1997{\natexlab{b}})}]{Koo_Interaction_1997}
Koo, B.-C. \& Moon, D.-S. 1997{\natexlab{b}}, ApJ, 485, 263

\bibitem[{Kristensen {et~al.}(2023)Kristensen, Godard, Guillard, Gusdorf, \&
  {Pineau des For{\^e}ts}}]{Kristensen_Shock_2023a}
Kristensen, L.~E., Godard, B., Guillard, P., Gusdorf, A., \& {Pineau des
  For{\^e}ts}, G. 2023, A\&A, 675, A86

\bibitem[{Lazendic {et~al.}(2010)Lazendic, Wardle, Whiteoak, Burton, \&
  Green}]{Lazendic_Multiwavelength_2010}
Lazendic, J.~S., Wardle, M., Whiteoak, J.~B., Burton, M.~G., \& Green, A.~J.
  2010, MNRAS, 409, 371

\bibitem[{Lee {et~al.}(2019)Lee, Koo, Lee, Burton, \&
  Ryder}]{Lee_infrared_2019}
Lee, Y.-H., Koo, B.-C., Lee, J.-J., Burton, M.~G., \& Ryder, S. 2019, ApJ, 157,
  123

\bibitem[{Lefloch {et~al.}(2012)Lefloch, Cabrit, Busquet, Codella, Ceccarelli,
  Cernicharo, Pardo, Benedettini, Lis, \& Nisini}]{Lefloch_CHESS_2012}
Lefloch, B., Cabrit, S., Busquet, G., {et~al.} 2012, ApJ, 757, L25

\bibitem[{Liszt \& Lucas(2001)}]{Liszt_Comparative_2001}
Liszt, H. \& Lucas, R. 2001, A\&A, 370, 576

\bibitem[{Liszt \& Lucas(1998)}]{Liszt_CO_1998}
Liszt, H.~S. \& Lucas, R. 1998, A\&A, 339, 561

\bibitem[{Liszt {et~al.}(2006)Liszt, Lucas, \& Pety}]{Liszt_Comparative_2006}
Liszt, H.~S., Lucas, R., \& Pety, J. 2006, A\&A, 448, 253

\bibitem[{Liu {et~al.}(2021)Liu, Wu, Zhang, Chen, Lin, Qin, Liu, Henkel, Wang,
  Liu, Yuan, Yuan, Li, Shen, Li, Esimbek, Wang, Li, Kim, Zhu, Madones,
  {Inostroza-Pino}, Meng, Zhang, Tatematsu, Xu, Ju, Kraus, \&
  Xu}]{Liu_Search_2021}
Liu, X.~C., Wu, Y., Zhang, C., {et~al.} 2021, ApJ, 912, 148

\bibitem[{Lucas \& Liszt(1996)}]{Lucas_Plateau_1996}
Lucas, R. \& Liszt, H. 1996, A\&A, 307, 237

\bibitem[{Lucas \& Liszt(2000)}]{Lucas_Comparative_2000}
Lucas, R. \& Liszt, H.~S. 2000, A\&A, 358, 1069

\bibitem[{Mangum \& Shirley(2015)}]{Mangum_How_2015}
Mangum, J.~G. \& Shirley, Y.~L. 2015, PASA, 127, 266

\bibitem[{Mazumdar {et~al.}(2022)Mazumdar, Tram, Wyrowski, Menten, \&
  Tang}]{Mazumdar_Submillimeter_2022a}
Mazumdar, P., Tram, L.~N., Wyrowski, F., Menten, K.~M., \& Tang, X. 2022, A\&A,
  668, A180

\bibitem[{Mendoza {et~al.}(2018)Mendoza, Lefloch, Ceccarelli, Kahane, Jaber,
  Podio, Benedettini, Codella, Viti, {Jimenez-Serra}, Lepine,
  {Boechat-Roberty}, \& Bachiller}]{Mendoza_search_2018}
Mendoza, E., Lefloch, B., Ceccarelli, C., {et~al.} 2018, MNRAS, 475, 5501

\bibitem[{Nakajima {et~al.}(2023)Nakajima, Takano, Tosaki, Taniguchi, Harada,
  Saito, Imanishi, Nishimura, Izumi, Tamura, Kohno, \&
  Herbst}]{Nakajima_Molecular_2023}
Nakajima, T., Takano, S., Tosaki, T., {et~al.} 2023, ApJ, 955, 27

\bibitem[{Neufeld \& Dalgarno(1989)}]{Neufeld_Fast_1989}
Neufeld, D.~A. \& Dalgarno, A. 1989, ApJ, 340, 869

\bibitem[{Park {et~al.}(2023)Park, Currie, Thomas, Rosolowsky, Dempsey, Kim,
  Rigby, Su, Eden, Colombo, Parsons, \& Moore}]{Park_12CO_2023}
Park, G., Currie, M.~J., Thomas, H.~S., {et~al.} 2023, ApJS, 264, 16

\bibitem[{Park {et~al.}(2013)Park, Koo, Gibson, h.~Kang, Lane, Douglas, Peek,
  Korpela, Heiles, \& Newton}]{Park_Shells_2013}
Park, G., Koo, B.~C., Gibson, S.~J., {et~al.} 2013, ApJ, 777, 14

\bibitem[{Park {et~al.}(2006)Park, Wakelam, \& Herbst}]{Park_Modeling_2006}
Park, I.~H., Wakelam, V., \& Herbst, E. 2006, A\&A, 449, 631

\bibitem[{Parravano {et~al.}(2003)Parravano, Hollenbach, \&
  McKee}]{Parravano_Time_2003}
Parravano, A., Hollenbach, D.~J., \& McKee, C.~F. 2003, ApJ, 584, 797

\bibitem[{Rho {et~al.}(2015)Rho, Hewitt, Boogert, Kaufman, \&
  Gusdorf}]{Rho_Detection_2015}
Rho, J., Hewitt, J.~W., Boogert, A., Kaufman, M., \& Gusdorf, A. 2015, ApJ,
  812, 44

\bibitem[{Rho {et~al.}(2001)Rho, Jarrett, Cutri, \&
  Reach}]{Rho_Near-Infrared_2001}
Rho, J., Jarrett, T.~H., Cutri, R.~M., \& Reach, W.~T. 2001, ApJ, 547, 885

\bibitem[{Sakai \& Yamamoto(2013)}]{Sakai_Warm_2013}
Sakai, N. \& Yamamoto, S. 2013, ChRv, 113, 8981

\bibitem[{Sato {et~al.}(1994)Sato, Mizuno, Nagahama, Onishi, Yonekura, \&
  Fukui}]{Sato_New_1994}
Sato, F., Mizuno, A., Nagahama, T., {et~al.} 1994, ApJ, 435, 279

\bibitem[{Shimajiri {et~al.}(2015)Shimajiri, Sakai, Kitamura, Tsukagoshi,
  Saito, Nakamura, Momose, Takakuwa, Yamaguchi, Sakai, Yamamoto, \&
  Kawabe}]{Shimajiri_Spectral-line_2015}
Shimajiri, Y., Sakai, T., Kitamura, Y., {et~al.} 2015, ApJS, 221, 31

\bibitem[{Shingledecker {et~al.}(2016)Shingledecker, Bergner, Le~Gal,
  {\"O}berg, Hincelin, \& Herbst}]{Shingledecker_Inference_2016}
Shingledecker, C.~N., Bergner, J.~B., Le~Gal, R., {et~al.} 2016, ApJ, 830, 151

\bibitem[{Shinn {et~al.}(2012)Shinn, Lee, \& Moon}]{Shinn_Ortho-to-para_2012}
Shinn, J.-H., Lee, H.-G., \& Moon, D.-S. 2012, ApJ, 759, 34

\bibitem[{Shirley(2015)}]{Shirley_Critical_2015}
Shirley, Y.~L. 2015, PASP, 127, 299

\bibitem[{Snow \& McCall(2006)}]{Snow_Diffuse_2006}
Snow, T.~P. \& McCall, B.~J. 2006, ARA\&A, 44, 367

\bibitem[{Stil {et~al.}(2006)Stil, Taylor, Dickey, Kavars, Martin, Rothwell,
  Boothroyd, Lockman, \& {McClure-Griffiths}}]{Stil_VLA_2006}
Stil, J.~M., Taylor, A.~R., Dickey, J.~M., {et~al.} 2006, AJ, 132, 1158

\bibitem[{Suzuki {et~al.}(1992)Suzuki, Yamamoto, Ohishi, Kaifu, Ishikawa,
  Hirahara, \& Takano}]{Suzuki_Survey_1992}
Suzuki, H., Yamamoto, S., Ohishi, M., {et~al.} 1992, ApJ, 392, 551

\bibitem[{Taniguchi {et~al.}(2024)Taniguchi, Gorai, \&
  Tan}]{Taniguchi_Carbon-chain_2024}
Taniguchi, K., Gorai, P., \& Tan, J.~C. 2024, Ap\&SS, 369, 34

\bibitem[{Taniguchi {et~al.}(2019)Taniguchi, Herbst, Ozeki, \&
  Saito}]{Taniguchi_Investigation_2019}
Taniguchi, K., Herbst, E., Ozeki, H., \& Saito, M. 2019, ApJ, 884, 167

\bibitem[{Tu {et~al.}(2024{\natexlab{a}})Tu, Chen, Zhou, {Safi-Harb}, \&
  Liu}]{Tu_Shock_2024}
Tu, T.-y., Chen, Y., Zhou, P., {Safi-Harb}, S., \& Liu, Q.-C.
  2024{\natexlab{a}}, ApJ, 966, 178

\bibitem[{Tu {et~al.}(2024{\natexlab{b}})Tu, Rayalacheruvu, Majumdar, Chen,
  Zhou, \& {Santander-Garc{\'i}a}}]{Tu_Yebes_2024}
Tu, T.-Y., Rayalacheruvu, P., Majumdar, L., {et~al.} 2024{\natexlab{b}}, ApJ,
  974, 262

\bibitem[{Tychoniec {et~al.}(2021)Tychoniec, {van Dishoeck}, {van't Hoff}, {van
  Gelder}, Tabone, Chen, Harsono, Hull, Hogerheijde, Murillo, \&
  Tobin}]{Tychoniec_Which_2021}
Tychoniec, {\L}., {van Dishoeck}, E.~F., {van't Hoff}, M. L.~R., {et~al.} 2021,
  A\&A, 655, A65

\bibitem[{Urquhart {et~al.}(2018)Urquhart, K{\"o}nig, Giannetti, Leurini,
  Moore, Eden, Pillai, Thompson, Braiding, Burton, Csengeri, Dempsey, Figura,
  Froebrich, Menten, Schuller, Smith, \& Wyrowski}]{Urquhart_ATLASGAL_2018}
Urquhart, J.~S., K{\"o}nig, C., Giannetti, A., {et~al.} 2018, MNRAS, 473, 1059

\bibitem[{{van der Tak} {et~al.}(2007){van der Tak}, Black, Sch{\"o}ier,
  Jansen, \& {van Dishoeck}}]{vanderTak_computer_2007}
{van der Tak}, F. F.~S., Black, J.~H., Sch{\"o}ier, F.~L., Jansen, D.~J., \&
  {van Dishoeck}, E.~F. 2007, A\&A, 468, 627

\bibitem[{{van Dishoeck} {et~al.}(1993){van Dishoeck}, Jansen, \&
  Phillips}]{vanDishoeck_Submillimeter_1993}
{van Dishoeck}, E.~F., Jansen, D.~J., \& Phillips, T.~G. 1993, A\&A, 279, 541

\bibitem[{Viti {et~al.}(2014)Viti, {Garc{\'i}a-Burillo}, Fuente, Hunt, Usero,
  Henkel, Eckart, Martin, Spaans, Muller, Combes, Krips, Schinnerer, Casasola,
  Costagliola, Marquez, Planesas, {van der Werf}, Aalto, Baker, Boone, \&
  Tacconi}]{Viti_Molecular_2014}
Viti, S., {Garc{\'i}a-Burillo}, S., Fuente, A., {et~al.} 2014, A\&A, 570, A28

\bibitem[{Wang \& Scoville(1992)}]{Wang_Strongly_1992}
Wang, Z. \& Scoville, N.~Z. 1992, ApJ, 386, 158

\bibitem[{Wolfire {et~al.}(2022)Wolfire, Vallini, \&
  Chevance}]{Wolfire_Photodissociation_2022}
Wolfire, M.~G., Vallini, L., \& Chevance, M. 2022, ARA\&A, 60, 247

\bibitem[{Yamagishi {et~al.}(2023)Yamagishi, Furuya, Sano, Izumi, Takekoshi,
  Kaneda, Nakanishi, \& Shimonishi}]{Yamagishi_Cosmic-ray-driven_2023}
Yamagishi, M., Furuya, K., Sano, H., {et~al.} 2023, PASJ, 75, 883

\bibitem[{Zhou {et~al.}(2022)Zhou, Zhang, Zhou, Arias, Koo, Vink, Zhang, Sun,
  Du, Zhu, Chen, Bovino, \& Lee}]{Zhou_Unusually_2022b}
Zhou, P., Zhang, G.-Y., Zhou, X., {et~al.} 2022, ApJ, 931, 144

\end{thebibliography}
\bibliographystyle{aa}

\begin{appendix} 
\onecolumn

\section{Spectra of all other detected transitions} \label{appendix}
\begin{figure*}[!h]
\centering
\includegraphics[width=0.99\textwidth]{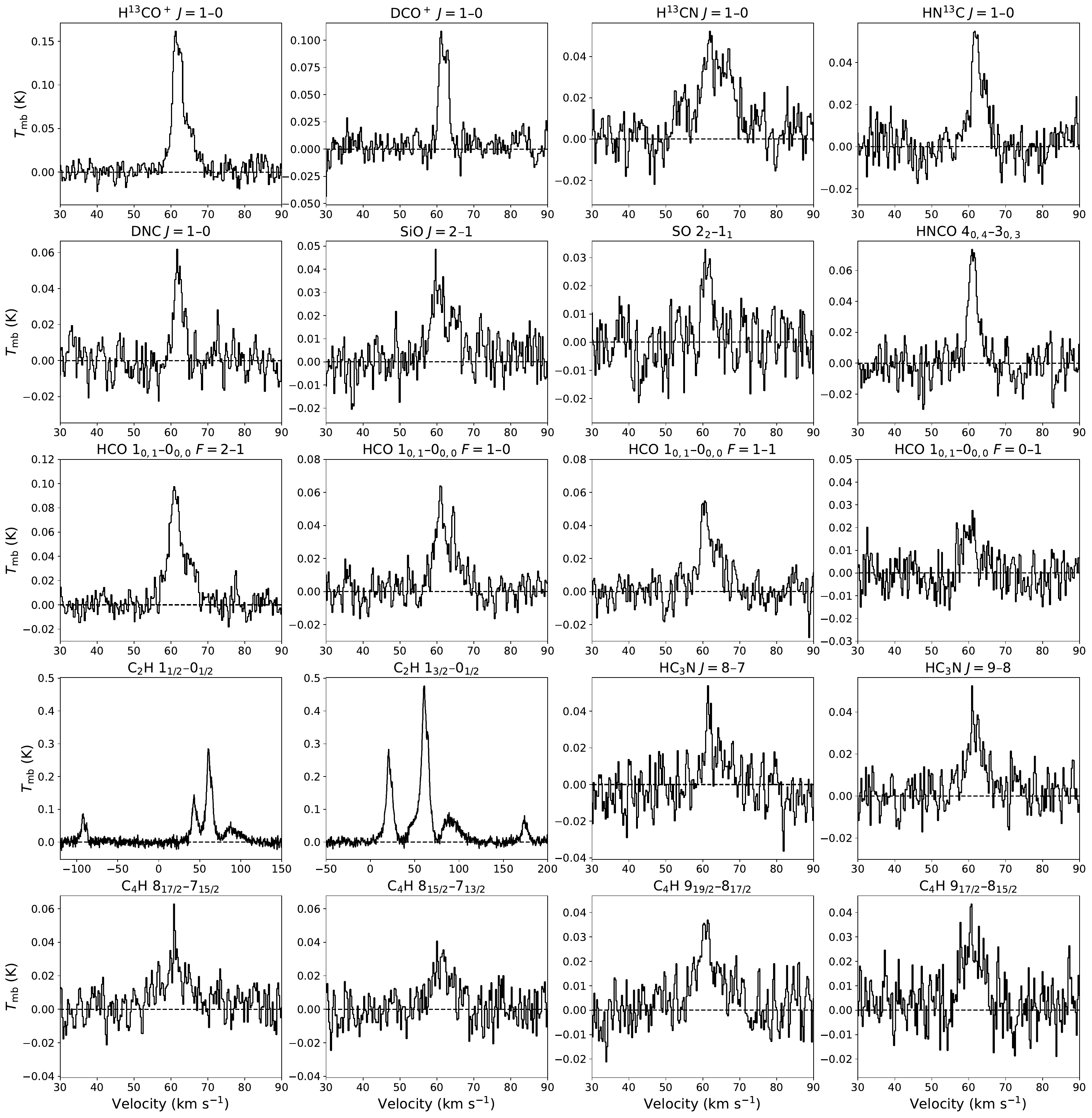}
\caption{Spectra of all detected molecular transitions averaged in the entire field of view except those which have been shown in Figure \ref{fig:spec}. 
}
\end{figure*}

\begin{figure*}[!h]
\centering
\includegraphics[width=0.99\textwidth]{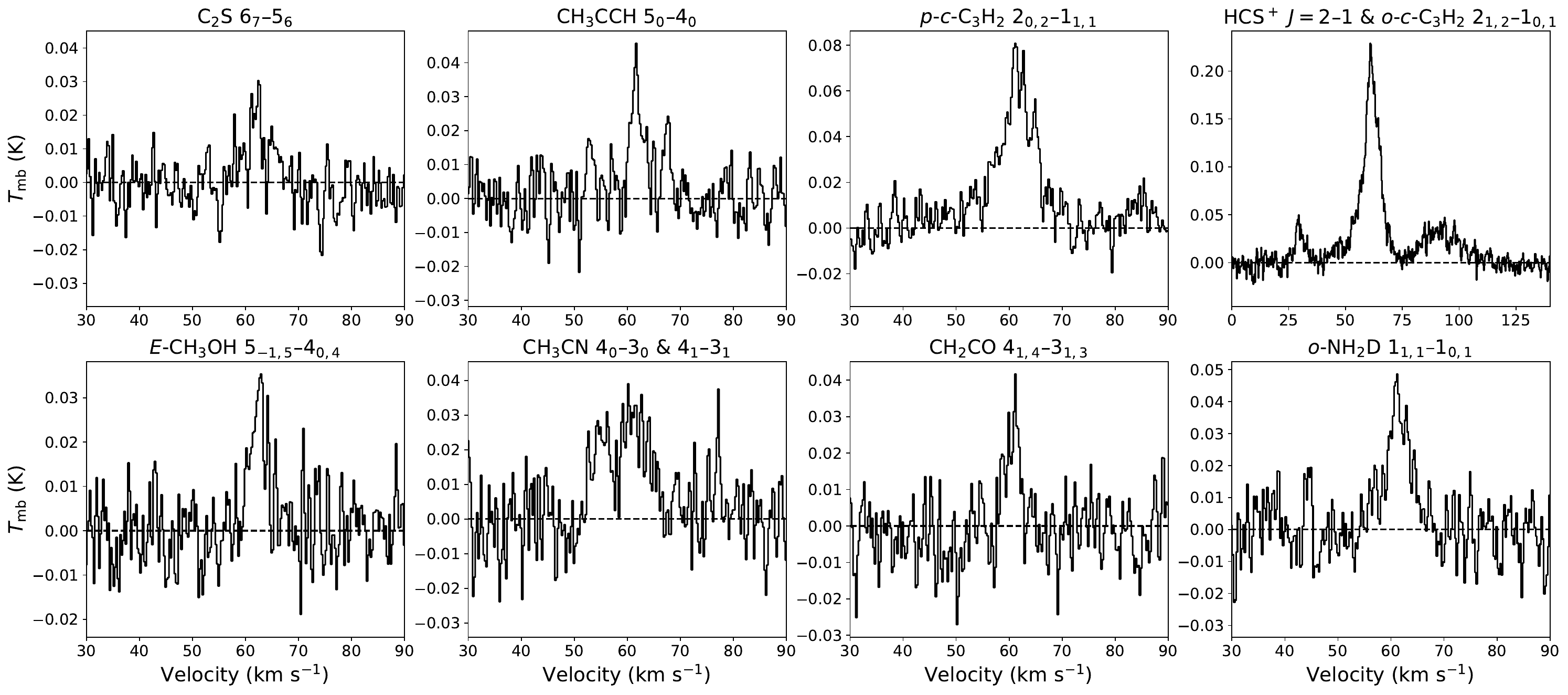}
\caption{ \textit{Continued.}
The $\rm HCS^+$ $J=2$--1 line is at $\approx +29$ \kms\ in the upper right panel of which the rest frequency is set to be the frequency of the \occthreehtwo\ line. 
}
\end{figure*}

\end{appendix}

\end{document}